\documentclass[11pt]{article}

\usepackage{graphicx}
\usepackage{epsfig}
\usepackage{bbm}
\usepackage{amsmath, amssymb}
\usepackage{cite}
\usepackage{subfig}
\usepackage{float}
\usepackage{braket}
\usepackage{resizegather}
\usepackage{multirow}
\usepackage{tikz}
\usepackage{longtable}
\usepackage{algorithm2e}

\usepackage[margin=2.5cm]{geometry}
\numberwithin{equation}{section}
\usepackage{hyperref}

\usepackage{color}

\newcommand{\IP}{\mathbb{P}}

\newcommand{\IZ}{\mathbb{Z}}
\newcommand{\cO}{{\cal O}}
\newcommand{\cN}{{\cal N}}

\newcommand{\cA}{{\cal A}}
\newcommand{\cB}{{\cal B}}
\newcommand{\cC}{{\cal C}}

\newcommand{\cV}{{\cal V}}

\newcommand{\npop}{N_{\rm{pop}}}

\def\cjn1{{\cA, \cC^*\otimes \wedge^j \cN^*}}
\def\bjn1{{\cA, \cB^*\otimes \wedge^j \cN^*}}
\def\vjn1{{\cA, \cV^*\otimes \wedge^j \cN^*}}
\def\cjn2{{\cA, \cC\otimes \wedge^j \cN^*}}
\def\bjn2{{\cA, \cB\otimes \wedge^j \cN^*}}
\def\vjn2{{\cA, \cV\otimes \wedge^j \cN^*}}

\newcommand{\varstr}[2]{\vrule height #1 depth #2 width0pt}

\begin{document}
\begin{flushright}
$ $\\
\end{flushright}
\vspace{8mm}
\begin{center}
{
\Large {\bf Evolving Heterotic Gauge Backgrounds: \\Genetic Algorithms versus Reinforcement Learning}\\[12pt]
\vspace{5mm}
\normalsize
{\bf{Steven Abel}$^{a,}$\footnote{s.a.abel@durham.ac.uk}},  
{\bf{Andrei Constantin}$^{b,}$\footnote{andrei.constantin@physics.ox.ac.uk}},  
{\bf{Thomas R.~Harvey}$^{b,}$\footnote{thomas.harvey@physics.ox.ac.uk}},
{\bf{Andre Lukas}$^{b,}$\footnote{andre.lukas@physics.ox.ac.uk}},
\bigskip}\\[0pt]
\vspace{0.23cm}
{\it 
{}$^a$IPPP, Durham University, Durham DH1 3LE, UK}\\[2ex]
{\it 
{}$^b$Rudolf Peierls Centre for Theoretical Physics, University of Oxford\\
Parks Road, Oxford OX1 3PU, UK
}\\[2ex]
\end{center}
\vspace{0.5cm}

\begin{abstract}\noindent
The immensity of the string landscape and the difficulty of identifying solutions that match the observed features of particle physics have raised serious questions about the predictive power of string theory. Modern methods of optimisation and search can, however, significantly improve the prospects of constructing the standard model in string theory. In this paper we scrutinise a corner of the heterotic string landscape consisting of compactifications on Calabi-Yau three-folds with monad bundles and show that genetic algorithms can be successfully used to generate anomaly-free supersymmetric $SO(10)$ GUTs with three families of fermions that have the right ingredients to accommodate the standard model. We compare this method with reinforcement learning and find that the two methods have similar efficacy but somewhat complementary characteristics.

\end{abstract}

\thispagestyle{empty}
\setcounter{page}{0}
\setcounter{footnote}{0}
\setcounter{tocdepth}{2}
\newpage

\tableofcontents

\newpage
\section{Introduction}
Obtaining string theory solutions that match the properties of particle physics in our world has been one of the main driving forces in string phenomenology, fuelling a significant effort in developing geometric engineering tools for constructing physically viable string models. This effort has met with only partial success due to the sheer mathematical difficulty of the problem and, to date, none of the existing string theory models allows for a proper description of particle physics. However, the model building experience of the past few decades has taught us much about the magnitude of the problem and about how (and also about how not) to approach it.\\[2mm]
The first lesson we have learnt is that the size of the string landscape is much larger than previously thought. The famous first estimate of $O(10^{500})$ consistent type IIB flux compactifications \cite{Douglas:2003um, Ashok:2003gk} seems rather conservative in comparison with the latest estimates. For instance, in Ref.~\cite{Taylor:2015xtz} it was shown that a single elliptically fibered four-fold gives rise to $O(10^{272,000})$ F-theory flux compactifications. The second lesson is that the number of compactifications that match the symmetry group and the particle spectrum of the Standard Model is very large, despite representing only a tiny fraction of all consistent compactifications to four dimensions. In Ref.~\cite{Constantin:2018xkj} it was argued that there are at least $10^{23}$ and very likely up to $10^{723}$ heterotic MSSMs, while the authors of Ref.~\cite{Cvetic:2019gnh} argued for the existence of a quadrillion standard models from F-theory. \\[2mm]
The third lesson is that these numbers are so large that traditional scanning methods cannot be used for systematic exploration. One can, of course, focus on small, accessible corners of the string landscape and this approach has been successful to some extent. For instance, in Refs.~\cite{Anderson:2013xka, Constantin:2018xkj}, $10^7$ pairs of Calabi-Yau three-folds and holomorphic bundles leading to $SU(5)$ heterotic string models that can accommodate the correct MSSM spectrum were explicitly constructed. However, going beyond just the spectrum is both computationally challenging and may well lead to an empty set of models, depending on the required degree of similarity with the standard model. What is needed instead is a tool-set of non-systematic search methods, that can quickly detect phenomenologically rich patches of the string landscape and which can implement `on the fly' a number of checks that go beyond the usual spectrum considerations.  \\[2mm]
Such methods can be developed using modern techniques of optimisation and search, and have already been implemented in a number of string contexts. Concretely, genetic algorithms (GAs) were introduced as a search method for finding string vacua
with viable phenomenological properties in Ref.~\cite{Abel:2014xta}, with initial applications in the  free fermionic formulation of heterotic string theory. More recently, reinforcement learning (RL) has been used in Ref.~\cite{Halverson:2019tkf} to generate type IIA intersecting brane configurations that lead to standard-like models and in Refs.~\cite{Larfors:2020ugo, Constantin:2021for} to identify $SU(5)$ and $SO(10)$ string GUT models, respectively, that can lead to three generation MSSMs. The landscape of type IIB flux vacua was  explored in Ref.~\cite{Cole:2019enn} using GAs and Markov chain Monte Carlo methods, while in Ref.~\cite{Krippendorf:2021uxu} the same methods were used, as well as RL. \\[2mm]
In the present paper we use a GA to obtain phenomenologically viable heterotic string compactifications on two Calabi-Yau three-folds with monad bundles. The same class of models has recently been explored in Ref.~\cite{Constantin:2021for} using RL, and indeed one of the purposes of the present work is to compare the two methods. We search and obtain several hundred $SO(10)$ models that can accommodate three standard model families of quarks and leptons. We also check during the course of the genetic evolution that the models are anomaly free and that they pass a necessary condition for bundle stability. The latter check would normally be performed at the end of the search, but recent advances in establishing simple, analytic formulae for line bundle cohomology \cite{Constantin:2018hvl, Klaewer:2018sfl, Larfors:2019sie, Brodie:2019dfx,Brodie:2019pnz,Brodie:2019ozt, Brodie:2020wkd, Brodie:2020fiq} have made possible the inclusion of additional checks during the search. \\[2mm]
\\[2mm]
The rest of the paper is organised as follows. In Section~\ref{sec:GA} we give a brief introduction to GAs while in Section~\ref{sec:heterotic_intro} we review the construction of heterotic models on Calabi-Yau three-folds with monad bundles. In Section~4 we discuss the details of our GA implementation and in Section 5 we present the results.

\section{The GA  technique for seeking perfect models}
\label{sec:GA}
Let us begin by giving some background and motivation for the GA. The approach of GAs (and  evolutionary computation more generally)  is to use concepts that are generally understood to underly evolution in order to find an optimum solution to a specified problem. Thus all GAs contain an evolving population of individual solution attempts, a ``{\it fitness}'' function for each individual, and a ``selection and breeding process'' based on their fitnesses, which applies evolutionary pressure with the aim of driving the population towards a solution. Depending on the problem at hand, success may be declared when the population contains sufficiently many ``good'' examples, or when one or several individuals in the population satisfies every one of a discrete set of desiderata. The present study will be of the latter kind. More specifically, we will be searching for models with properties that make them candidates for a standard model from string theory. For easy of terminology we will refer to such models as ``perfect models"  in the following.\\[2mm]
The idea of evolutionary computing was developed in the 1950s, but it was formalised by Holland \cite{Holland1975} who moulded the idea of ``building blocks'' into a theory of ``schema''. Later reviews can be found in Refs.~\cite{David1989,Holland1992,Jonesarticle,Reeves2002,Charbonneau:2002,haupt,Michalewicz2004,MCCALL2005205,Whitearticle}. There has been growing interest in the technique within the astrophysics and particle theory community, for example in Refs.~\cite{Charbonneau:1995,Yamaguchi:1999hq,Metcalfe:2002pd,Allanach:2004my,Akrami:2009hp,Nesseris:2012tt,
Blaback:2013ht,Damian:2013dq, Damian:2013dwa,Blaback:2013fca,Blaback:2013qza,
Abel:2014xta,Hogan:2014qsa,Behr:2015oqq,Ruehle:2017mzq, Abel:2018ekz,Bull:2018uow,Cole:2019enn,Cole:2020ktw,AbdusSalam:2020ywo,CaboBizet:2020cse,Bena:2021wyr,Faraggi:2021yjz}, partly due to its often surprising efficacy.\\[2mm]
We will return to discuss some of the ideas that have been proposed to explain the efficacy of GAs in a moment, but let us first describe the technique itself in more detail as it pertains to the  problem under discussion. The fundamental starting point of any GA is to encode the problem into a string of data, referred to as the {\it genotype} (each entry of which is referred to as an {\it allele}). \\[2mm]
To be specific, in the present case the genotype is constructed as follows. Fixing a Calabi-Yau three-fold $X$, we seek monad bundles $V$ constructed from two sums of holomorphic line bundles $B$ and $C$, with ranks $r_B$ and $r_C$, respectively, via the short exact sequence $0 \rightarrow V \rightarrow B \rightarrow C \rightarrow 0$. Each line bundle in $B$ and~$C$ is specified by its first Chern class, which, relative to a basis of $H^2(X)$, is described by an integer list with length equal to the Picard number $h=h^{1,1}(X)$ of $X$. This means that $V$ is specified by $h(r_B+r_C)$ integers. In fact, as explained below, we will be looking for bundles satisfying $c_1(V)=0$, a constraint that cuts down the number of integer parameters to $h(r_B+r_C-1)$. In order to translate this into a genotype we will allow the line bundle integers to assume values only within a finite range, 
which can then be implemented as a binary encoding, with each integer being mapped to a binary string of fixed length $n_B$ for $B$ and $n_C$ for $C$.  
 \\[2mm]
Thus the collection of all line bundle integers in $B$ and $C$ or, rather, the corresponding bit-string obtained by joining their binary encodings defines a particular model. The starting point of the GA is then a population of $\npop$ randomised individuals, each being represented as a string of digit alleles of length 
\begin{equation}
\ell ~=~ h (r_Bn_B+(r_C-1)n_C)~.
\end{equation}
For instance, for the first of our working examples we will have $h=2$, $n_B=n_C=3$, $r_B=6$ and $r_C=2$, giving $\ell=42$. \\[2mm]
In the case of such a binary encoding there are general arguments \cite{David1989,Reeves2002,Michalewicz2004}
that suggest that the optimum size of the population $\npop$ depends only logarithmically on the genotype length $\ell$, with a formal lower bound 
\begin{equation}
\label{eq:npop}
\npop ~\gtrsim~ 1+\frac{\log(-\ell /\log P_* )}{\log 2}~,
\end{equation}
determined by requiring that both possible values for each allele should appear 
in the initially randomised population with some 
representative probability $P_*$. (For example $P_*=0.999$, would imply that there is a 1 in $10^3$ probability that any individual allele contains only a 0 or a 1 in the entire population.) Typically this gives an estimate of $\npop\gtrsim 20$ that is a significant underestimate of 
the optimal size of the population, but it nevertheless gives an indication of the required population size. For the current study, the GA was found to perform well with $\npop\gtrsim 100$. The GA then consists of repeated application of the following  basic elements on the population: \\[2mm]
{\bf Selection:} Individuals are first selected from the population to make $\npop$ ``breeding pairs'' which will create a new generation to replace the original one. In order to do this the first step is to calculate the properties of each individual (its so-called {\it phenotype}). This particular step of extracting the phenotypes from all the genotypes is the true hallmark of a GA, as the relation between the genotype and the phenotype it generates, and also the correlations between different properties within the phenotype, may be very deeply hidden. This step, which is where the physics comes into play, is  also often the most case-dependent and computationally-intensive part of the whole procedure. Encouragingly, however, it is the aspect that is most amenable to parallelisation.\\[2mm]
Based on the phenotype, each individual is then assigned  a fitness value, which is a function of the individual's phenotype  designed such that it increases as the model approaches a desired optimal solution. There is a wide variety of methods for assigning the fitness to an individual, and the process may take many different forms, such as rank-weighting, tournament selection, and so on. Designing a good fitness function is a crucial factor in the success (or otherwise) of the GA, because its convergence will depend strongly on  the correlation between the fitness and the distance from the hypothetical optimal solution in genotype space (that is, the Hamming distance). This fitness-distance-correlation was formalised in a number of works by Jones, Forrest {\it et al} \cite{Forrest1993,Jonesarticle,Jones1995,RePEc:wop:safiwp:95-02-022,Collard1998}. \\[2mm]
We can appreciate this aspect of the GA for the simple example of maximising a smooth function. Suppose we choose a fitness functions which is simply proportional to the function value. Then, as the population approaches a relative flat maximum, there may be little difference in fitness between individuals which are nevertheless widely separated in coordinate space (which in the case of a continuous function would be directly given by the genotype). In other words there would be little fitness-distance-correlation in the neighbourhood of a solution and convergence would cease. A simple way to avoid this kind of convergence problem is to rank the individuals in the population by their fitness value, and then to simply equate their fitness with their ranking. This is the aforementioned rank-weighting selection procedure, and is the approach that we will adopt in this study. It has the additional advantage that we do not  need to store all fitness values separately, but can keep track of the ranking by re-ordering all the $\npop$ genotypes within a single $\npop\times\ell$ array. Selection to participate in breeding pairs is then carried out with a linearly distributed likelihood (known generally as roulette wheel selection) that simply depends on an individual's ranking. In the present case an individual at rank $k$ in the population is selected for breeding with a probability that depends linearly on its ranking as
\begin{equation}\label{eq:Pk}
P_k ~=~ \frac{2}{(1+\alpha) \npop }  \left( 1+\frac{\npop-k } {\npop-1} (\alpha-1)\right)~,
\end{equation} 
where $\alpha $ determines how much more frequently the fittest individual is chosen over the least fit. In other words, the probabilities $P_{\npop}$ and $P_1$ for selecting the fittest and least fittest individual are related by
\begin{equation}
P_1 ~=~  \frac{2\alpha}{(1+\alpha)\npop} ~=~ \alpha P_{\npop} ~. 
\end{equation} 
Typically, an optimal configuration is found when the fittest individual is selected for breeding a few times more than the least fit, so that a good choice would be $\alpha = 2$ or 3. But note that it is important that less fit individuals are also able to breed. \\[2mm]
{\bf Breeding/Cross-over:}  
At this stage we have $\npop$ breeding pairs, with the median individual having been selected for breeding twice (that is, $\npop P_{\npop/2}= 1+{\cal O}(1/\npop)$).  A new population of individuals is then formed by splicing together the genotypes of the two individuals in each breeding pair. Again there are many different ways to do this. The canonical choice is $N$-point cross-over, namely one cuts the genotypes of a breeding pair at the same $N$ random positions along their lengths and swops the cut sections. For robustness, the publicly available codes often use sophisticated combinations (for example \cite{Charbonneau:1995,Charbonneau:1995b,Charbonneau:2002,Charbonneau:2002b} uses both one- and two-point cross-over in roughly equal proportions) to reduce end-point biasing. In the present study this effect was not found to be significant, and a simple one-point cross over was found to work sufficiently well.\\[2mm]
{\bf Mutation:} With only the two previous elements, one would already observe clustering of the population around good solutions over generations. However, the  power of GAs comes from the third element which is {\it mutation}. Once a new generation is formed, a small fraction (usually around a percent) of the  alleles are flipped at random. This prevents stagnation in the population, where the entire population clusters around a local maximum in the fitness, when better solutions are available globally. For example, it removes the possibility alluded to below Eq.~\eqref{eq:npop} that a particular allele could end up being blocked with either all 0's or all 1's running throughout the entire population. 

It is important to understand that mutation is not just an improvement to the convergence, but is absolutely integral to the entire process. Depending on the problem and the structure of the fitness landscape, it causes a dramatic increase in the overall rate of convergence. This can be seen in practice by optimising the mutation rate as demonstrated  in  Ref.~\cite{Abel:2014xta}. It is worth mentioning one possible innovation that could be considered, but which we did not use in this study, namely so-called {\emph {creep mutation}}. This is effectively a small mutation in the {\it phenotype} which overcomes so-called Hamming walls, which occur when the population is close to an optimum solution in terms of phenotype, but far away in terms of Hamming distance. An example would be the binary number 1000 and 0111 which are numerically close but have four different digits. This can again be thought of as a way to overcome breakdowns in the fitness-distance-correlation, and could be thought of as {\it assisted mutation}.\\[2mm]
{\bf Elitism:}
Finally to create the new generation in a way that guarantees a monotonically increasing maximum fitness, we invoke {\it elitism}. This means the fittest individual from the previous generation is copied into the new one and replaces the least fit new individual.\\[2mm]
The process is then repeated over many generations, and is terminated and reset when each ``perfect model'' is found. Note that there are many other practical elements that could be considered, such as fitness ``crowding penalties'', and  ``niching'', that we do not discuss (or require). They are covered in the literature (see Refs.~\cite{Reeves2002,haupt}). The overall procedure that we use in this study is summarised in the following pseudo-code: \\[2mm]

\begin{algorithm}[H]
 \KwResult{`Perfect' Standard Model  built on Calabi-Yau manifold}
 \For{population}{Initialize genotype}
 \While{generations $<$ gen-number}  
{
\For{population}{ 
 Find phenotype (model properties) \\ Rank individuals by fitness\\ Elitism: Copy fittest individual}
 {\For{new-population} {Rank-weighted roulette wheel selection of breeding pair from {\it population}\\ One-point crossover to create new individuals \\ 
 Mutate new individuals}
 }{Elitism: Paste previous top-ranked individual over bottom-ranked\\ {\it population} $=$ {\it new-population} }}
  \caption{\label{alg} \it Pseudocode for GA}
\end{algorithm}

\subsection{Discussion of efficacy} 
Before moving on to the details of heterotic string models with monad bundles, it is worth pausing to consider why GAs are thought to work. With the benefit of several intervening decades of development in heuristic search methods, we can see that GAs contain an interesting mix of elements  that can be found in other  methods. For example the fitness function is arguably just a cost-function by another name. In addition, the general idea of fitness-distance-correlation in Jones, Forrest {\it et al} \cite{Forrest1993,Jonesarticle,Jones1995,RePEc:wop:safiwp:95-02-022,Collard1998} clearly has some parallels with Gradient-Descent or Nelder-Mead search techniques. However, one can also see that this correlation can only be a part of the whole story and that GAs must confer their own advantages: indeed a very strong correlation would render the genotype representation superfluous because the dynamics of the flow could be mapped directly to the phenotype. Some power of GAs clearly lies then in the fact that small changes in the genotype can result in huge changes in the phenotype, so that the GA samples the whole space before it begins to converge. Holland's schema theory idealised this behaviour in terms of hidden building blocks within the genotype. A schema is a representation of some crucial set of digits that confers some favourable characteristic, and in this instance might
look something like $S=1***\,0\,0\,*$\,. This example has 3 entries that
we are interested in (hence we say it is order $3$) and 4 entries
that we do not care about, which are labelled with a wildcard $*$. The probabilistic arguments of Holland (summarised in \cite{Abel:2014xta}) 
then suggest that schemata of this kind are important because selection favours
the propagation of shorter strings of data: small subsections of the
genome that confer fitness dominate first and, once they are shared
by the majority of the population, crossover does not affect them. As a result, so the theory goes,  if the schema $S$ confers fitness on an individual, 
then it will grow exponentially within the population until an equilibrium is reached where the attraction to the solution balances against the repulsion due to the mutation. \\[2mm]
Thus the overall behaviour suggested by schema theory is that a GA incorporates and balances competing forces. Selection and breeding tends to produce convergence around local maxima in the fitness landscape, drawing the population in over generations. On the other hand the effect of mutation is to push the population away from local maxima (on average), so that as a whole it can explore the entire parameter space. The power of GAs is then that they are sensitive to the entire landscape, but simultaneously can respond to and converge on interesting regions. \\[2mm]
The schema argument fell temporarily out of favour in the 1990s, for reasons which are summarised in Ref.~\cite{Whitearticle},
 mainly on the grounds that it does not consider the constructive effects of cross-over, and that it assumes a constant contribution to the fitness of a given schema. In reality, the population is not static, so that changes in the rest of the genotype tend to  alter the relative fitness of the schema after the first few generations in a run. Despite this objection, the idea was rehabilitated by Poli {\it et al} in a suite of papers (see Ref.~\cite{Whitearticle} for a comprehensive review) which developed so-called {\it exact} schema based models, and showed their relation to Markov chains. Simultaneously there have been several other approaches to understanding GAs, for example statistical mechanical approaches \cite{PhysRevLett.72.1305} or stochastic differential equations \cite{heredia}, which each seem to have their own drawback. In the former approach, one assigns the fitness function with a Boltzmann weighting, while the latter approach is restricted to simple cases such as the so-called onemax problem which is equivalent to solving the one-dimensional Ising model. In these cases it is likely that the set-up of the configuration influences the results so that the conclusions are not general but really only applicable to the very specific configurations used to frame the argument. Despite these limitations, the schema theory remains as one of the best interpretations of the behaviour of the GA.

\section{Heterotic models with monad bundles}
\label{sec:heterotic_intro}
Having presented the main features of GAs, we now turn to a concise discussion of the geometric ingredients entering the construction of $E_8\times E_8$ heterotic string compactifications on Calabi-Yau three-folds with monad bundles.  More details can, for example, be found in Refs.~\cite{Distler:1987ee,Kachru:1995em,Anderson:2008ex,Anderson:2008uw,Anderson:2009mh,He:2009wi,Constantin:2021for}. In particular, we will specify the degree of similarity to the properties of the standard model, that qualifies a particular heterotic model to be designated as  ``perfect". From the perspective of the GA this amounts to working out the phenotype.

\subsection{Heterotic CY models}
As already alluded to, heterotic $E_8\times E_8$ compactifications are based on the data $(X,V,\tilde{V},C)$, where $X$ is a Calabi-Yau three-fold, $V$ and $\tilde{V}$ are slope poly-stable vector bundles over $X$ whose structure groups embed into $E_8$ and $C$ is an effective curve class on $X$. Poly-stability is the mathematical condition to ensure the bundles preserve supersymmetry and it is, in general, difficult and time-consuming to check explicitly. For the purpose of our algorithmic approach we will only check a necessary condition known as Hoppe's criterion.  It says that a poly-stable bundle $V$ over $X$ must necessarily satisfy
\begin{equation}\label{hoppe}
h^0(X,\wedge^r V) = 0\;,\quad \text{for all } r = 1, 2,\ldots, {\rm rk}(V)-1.
\end{equation}
Another constraint on the bundle $V$ arises from the fact that only special unitary groups can be embedded into $E_8$. This means that we require 
\begin{equation}
 c_1(V)=c_1(\tilde{V})=0\; .
\end{equation} 
{\bfseries Anomaly cancellation:} The sectors of the four-dimensional theory which originate from the bundle $\tilde{V}$ and the five branes wrapping a curve with class $C$ are hidden sectors, in the sense that they couple to the observable sector associated to $V$ only gravitationally. Since we are interested in the observable particle physics, these sectors will not be constructed explicitly. However, for a consistent model, we need to satisfy the anomaly cancellation condition
\begin{equation}
{\rm c}_2(TX)+{\rm ch}_2(V)+{\rm ch}_2(\tilde{V})=C
\end{equation}
which relates the various sectors of the theory. This is done by demanding that
\begin{equation}\label{eq:anomcond}
 {\rm c}_2(TX)+{\rm ch}_2(V)\in\mbox{Mori cone of }X
\end{equation}
which guarantees the existence of a supersymmetric completion of the model with a trivial hidden bundle $\tilde V$ and five-branes wrapping a curve with class $C= {\rm c}_2(TX)+{\rm ch}_2(V)$.\\[2mm]
{\bfseries Spectrum:} For the purpose of this paper, we will consider ``observable" bundles $V$ with structure group $SU(4)$ which breaks one of the $E_8$ symmetries to an $SO(10)$ GUT group. The $SO(10)$ multiplets arise from the decomposition of the ${\bf 248}$ adjoint of $E_8$ under the maximal subgroup $SU(4)\times SO(10)$, which reads
\begin{equation*}
\begin{array}{lcccccccccc}
\textbf{248}_{E_8} ~\rightarrow &(\textbf{1},\textbf{45}) &\oplus&(\textbf{4},\textbf{16})& \oplus\!\!\!\!\!&(\overline{\textbf{4}},\overline{\textbf{16}}) &\!\!\!\!\!\oplus&(\textbf{6},\textbf{10}) &\oplus&(\textbf{15},\textbf{1})&\\[2mm]
& \text{gauge}&&\text{families}&&\text{anti-families}&&\text{Higgs}&& \text{bundle}&\\[-2pt]
& \text{bosons}&& && &&&& \text{moduli}& \\[2mm]
&&& n_{\bf 16}=h^1(V)&&n_{\overline{\bf 16}}=h^1(V^*)&&n_{\bf 10}=h^1(\Lambda^2 V)&&n_{\bf 1}=h^1(V\otimes V^*)
\end{array}
\end{equation*}
The number of each multiplet is given by the first cohomology of $V$, $V^*$ and its various products, as indicated in the last row above. Of particular importance is the chiral asymmetry of families which can be expressed in terms of the index of $V$ as
\begin{equation}
	\label{eqIndex}
	n_{\mathbf{16}}- n_{\overline{\mathbf{16}}} = -\mbox{ind}(V) = h^1(X,V) - h^2(X,V)\; .
\end{equation}
The last equality in the above formula assumes that $h^0(V)=h^3(V)=0$ which follows from the poly-stability of $V$, via Hoppe's criterion~\eqref{hoppe}.\\[2mm]
{\bfseries Equivariance:} The $SO(10)$ GUT symmetry needs to be broken to the standard model group and this is done by including a discrete Wilson line into the construction, which is possible when the manifold is non-simply connected. However, most of the typical constructions lead to Calabi-Yau three-folds that are simply connected. On the other hand, if a freely-acting symmetry $\Gamma$ on $X$ exists, this can be used to produce a quotient $\hat{X}=X/\Gamma$ with non-trivial fundamental group. The Wilson line is then included on the resulting quotient manifold $\hat{X}=X/\Gamma$. In order to end up with three chiral families after performing this quotient, the upstairs model on $X$ requires a chiral asymmetry of
\begin{equation}\label{eq:families}
 n_{\mathbf{16}}- n_{\overline{\mathbf{16}}} = -\mbox{ind}(V)\stackrel{!}{=}3|\Gamma|\; ,
\end{equation}
where $|\Gamma|$ is the order of the discrete group $\Gamma$. For this construction to make sense, the upstairs bundle $V\rightarrow X$ needs to descend to a bundle $\hat{V}\rightarrow\hat{X}$ on the quotient CY and this is the case iff $V$ admits a $\Gamma$-equivariant structure. This is typically a non-trivial constraint which will be discussed in more detail below, when we introduce monad bundles.

\subsection{Calabi-Yau three-folds}
To realise the above construction we require, in a first instance, explicit CY manifolds. It turns out that, for group-theoretical reasons, the Wilson-line breaking of $SO(10)$ to the Standard Model requires a discrete group $\Gamma$ which is at least $\mathbb{Z}_3\times \mathbb{Z}_3$. While the number of known CY manifolds is large, only a handful of known examples have symmetry groups this large~\cite{Braun:2010vc,Braun:2017juz, Candelas:2016fdy, Constantin:2021for}, so the choice of manifold is, in fact, rather constrained~\footnote{This changes for models based on an $SU(5)$ GUT group whose breaking only requires a $\mathbb{Z}_2$ symmetry.}. For the purpose of this paper, we will focus on two of these manifolds, both realised as complete intersections in products of projective spaces (CICYs) and represented by the following configuration matrices
\begin{equation}\label{eq:CYs}
 \left[\begin{array}{c|c}\mathbb{P}^2&3\\\mathbb{P}^2&3\end{array}\right]^{2,83}_{-162}\;,\qquad
 \left[\begin{array}{c|ccc}\mathbb{P}^2&1&1&1\\\mathbb{P}^2&1&1&1\\\mathbb{P}^2&1&1&1\end{array}\right]^{3,48}_{-90}~.
\end{equation}
These manifolds are also known as the bi-cubic CY and the triple tri-linear CY, respectively. They are embedded in the ambient spaces ${\cal A}=(\mathbb{P}^2)^{\times 2}$ for the bi-cubic and ${\cal A}=(\mathbb{P}^2)^{\times 3}$ for the triple tri-linear. The manifold itself is cut out by the common zero locus of homogeneous polynomials whose multi-degrees correspond to the columns of the above matrices. The superscripts provide the Hodge numbers $h^{1,1}(X)$, $h^{2,1}(X)$ and the subscript the Euler number, $\eta(X)$, of the manifold. Both manifolds have a freely-acting symmetry $\Gamma=\mathbb{Z}_3\times\mathbb{Z}_3$, as required for the $SO(10)$ GUT breaking. We will also need the second Chern class $c_{2i}(TX)$ of the tangent bundles and the triple-intersection numbers $d_{ijk}$. Relative to the obvious basis  $(J_1,\ldots ,J_h)$ of $H^2(X)$, obtained by restricting the K\"ahler forms of the projective ambient space factors to $X$, and the associated dual basis of $H^4(X)$, they are given by
\begin{equation}
 \begin{array}{lll}
  \mbox{bi-cubic}\qquad&c_{2i}(TX)=(36,36)\qquad&d_{122}=d_{112}=3\\
  \mbox{triple tri-linear}&c_{2i}(TX)=(36,36,36)&d_{iij}=3\;\;\forall i\neq j\;,\quad d_{123}=6\; .
\end{array}
\end{equation}
  
\subsection{Monad bundles}
Explicit bundle construction frequently rely on holomorphic line bundles as basic building blocks. Holomorphic line bundles are classified by their first Chern class. Relative to an integral basis $(J_1,\ldots ,J_h)$ of $H^2(X)$, the first Chern class of a line bundle $L\rightarrow X$ can be represented by an integer vector ${\bf k}\in\mathbb{Z}^h$, such that $c_1(L)=k^iJ_i$, hence the notation $L={\cal O}_X({\bf k})$.\\[2mm]
Monad bundles are defined by a short exact sequence
\begin{equation}\label{MonadSeqence}
0 \longrightarrow V \longrightarrow B \stackrel{f}{\longrightarrow} C \longrightarrow 0\;,\qquad
B = \bigoplus_{i=1}^{r_B} \mathcal O_X(\mathbf b_i)\;,\quad C = \bigoplus_{a=1}^{r_C} \mathcal O_X(\mathbf c_a)\; ,
\end{equation}
where, for our purposes, $B$ and $C$ are each taken to be line bundle sums. Thus such monad bundles are specified by a pair $(B,C)$ of line bundle sums which can be identified with an integer matrix $({\bf b}_1,\ldots ,{\bf b}_{r_B},{\bf c}_1,\ldots ,{\bf c}_{r_C})$ of size $h\times (r_B+r_C)$. The bundle $V$ is isomorphic to the kernel, ${\rm Ker}(f)$ of the monad map $f$ which can be thought of as a $r_C\times r_B$ matrix with polynomial entries $f_{ai}\in\Gamma({\cal O}_X({\bf c}_a-{\bf b}_i))$.\\[2mm]
A given choice of $(B,C)$ might lead to a monad sheaf, rather than a monad bundle. This happens when the degeneracy locus of $f$, that is, the locus on $X$ where the rank of $f$ is less than maximal, is non-trivial. Since we are considering smooth, geometrical models we should avoid sheafs, so we have to check ``bundleness" of $V$ by computing the dimension
\begin{equation}\label{eq:deglocus}
 d_{\rm deg}=\mbox{dimension of degeneracy locus of }f\; .
\end{equation} 
Typically, the matrix $f$ exhibits a pattern of zero and non-zero entries, depending on whether $\Gamma({\cal O}_X({\bf c}_a-{\bf b}_i))$ is trivial or non-trivial, and for otherwise generic choices, the dimension $d_{\rm deg}$ can be determined from this pattern. The model is acceptable iff $d_{\rm deg}=-1$.\\[2mm]
A monad map with too many zero entries might also lead to a split bundle, that is, to a bundle $V$ whose structure group is a proper sub-group of $SU(4)$, for example $S(U(2)\times U(2))$. These split bundles lead to a GUT group larger than $SO(10)$ and we do not consider this possibility here. To this end, we introduce the number
\begin{equation}\label{eq:nsplits}
 n_{\rm split}=\mbox{number of splits of the bundle }V
\end{equation}
which can be determined from the pattern of the monad map. It is also useful to introduce the number
\begin{equation}\label{eq:nOX}
 n_{\rm trivial}=\mbox{number of trivial line bundles }{\cal O}_X\mbox{ in }(B,C)\; .
\end{equation} 
As mentioned earlier, $\Gamma$-equivariance of $V$ is a non-trivial constraint but a full check based on presently available methods is not compatible with an efficient computational realisation. Instead, we will perform a strong necessary check for $B$ and $C$ to possess a $\Gamma$-equivariant structure. This is done by checking that each unique line bundle $L$ in $B$ or $C$ has an index divisible by $|\Gamma|$. If a line bundle $L$ appears with multiplicity $m$ in either $B$ or $C$, we require that
\begin{equation}\label{eq:equivBC}
 {\rm ind}(L^{\oplus m})\quad\mbox{is divisible by}\quad |\Gamma|\; .
\end{equation} 
To check the other constrains we require the following formulae for the Chern classes of a monad bundle:
\begin{equation}\label{eq:monchern}
\begin{array}{rcl}
 {\rm rk}(V)&=&{\rm rk}(B)-{\rm rk}(C)\stackrel{!}{=}4\\
c_1^k(V)&=& \displaystyle c_1^k(B)-c_1^k(C)=\sum_{i=1}^{r_B}b_i^k- \sum_{a=1}^{r_C} c_a^k\stackrel{!}{=}0\\
 c_{2k}(V)&=&\displaystyle  {\rm ch}_{2k}(C)-{\rm ch}_{2k}(B)=\frac{1}{2}d_{klm}\left(\sum_{a=1}^{r_C}c_a^lc_a^m-\sum_{i=1}^{r_B}b_i^lb_i^m\right)\stackrel{!}{\leq}c_{2k}(TX)\\[4pt]
 {\rm ind}(V)&=& \displaystyle \sum_{q=0}^3(-1)^q h^q(X,V)=\frac{1}{2}c_3(V)={\rm ch}_3(B)-{\rm ch}_3(C)\\[4pt]
 &=& \displaystyle \frac{1}{6}d_{klm}\left(\sum_{i=1}^{r_B}b_i^kb_i^lb_i^m-\sum_{a=1}^{r_C}c_a^kc_a^lc_a^m\right)\stackrel{!}{=}-3|\Gamma|~.
\end{array} 
\end{equation}
The conditions appearing on the right-hand sides of these equations originate from the requirement that $V$ has an $SU(4)$ structure group, from the anomaly cancellation condition~\eqref{eq:anomcond} and from the constraint~\eqref{eq:families} on the chiral asymmetry.\\[2mm]
Calculating cohomology for monad bundles requires the long exact sequence in cohomology associated to the monad sequence~\eqref{MonadSeqence}. Such computations can require ranks of maps and can be time-consuming. For this reason, checking Hoppe's criterion for poly-stability in full is not feasible while running a GA. Instead, we will use that Hoppe's criterion is violated and, hence, that $V$ is non-supersymmetric if
\begin{equation}\label{eq:Hoppetest}
 h^0(B)-h^0(C)>0\quad\mbox{or}\quad h^0(B^*)-h^0(C^*)>0\; .
\end{equation}
Fortunately, analytic formulae for zeroth line bundle cohomology exist for both manifolds under considerations~\cite{Constantin:2021for},  so the above expressions can be checked efficiently. 

\section{Monad bundles and genetic algorithms}
\subsection{The environment}
Let us now outline the ``environment" that will form the basis of our search.  We will consider models consisting of monad bundles~\eqref{MonadSeqence}, specified by matrices $(B,C)$, on a fixed CY three-fold $X$, in practice one of the two manifolds in Eq.~\eqref{eq:CYs}. The integer entries in $(B,C)$ are not, a priori, bounded but will be restricted as
\begin{equation}
 b_{\rm min}\leq b_i^k\leq b_{\rm max}\;,\qquad c_{\rm min}\leq c_a^k\leq c_{\rm max}\; ,
\end{equation}
so that the environment becomes finite. Given that we are aiming to match relatively small integers, such as the required index, this does not seem to be a serious limitation. 
In fact, model building experience suggests~\cite{Anderson:2011ns,Anderson:2012yf,Anderson:2013xka,Constantin:2021for} that viable models typically arise for relatively small integer entries. The first Chern class constraint, $c_1(V)=0$, can be easily solved (see Eq.~\eqref{eq:monchern}), for example by fixing the last line bundle in $C$ in terms of the others. For this reason, we constrain the environment to pairs $(B,C)$ that satisfy $c_1(B)=c_1(C)$. If we allowed, say, $10$ values per entry, the number of states in the environment is of  order
\begin{equation}
 10^{h(r_B+r_C-1)}\; .
\end{equation}
Since $r_B+r_C-1\geq 5$ and we consider $h=h^{1,1}(X)\in\{2,3\}$ this is quite sizeable and not suited for systematic scanning. Model building experience also shows~\cite{Anderson:2011ns,Anderson:2012yf,Anderson:2013xka,Constantin:2021for} that the fraction of viable models within this environment is so small that a purely random search could not be successful.\\[2mm]
The environment has a large degeneracy, governed by the symmetry
\begin{equation}\label{eq:redsymm}
 H\times S_{r_B}\times S_{r_C}\; ,
\end{equation} 
where the first factor is $H=S_2$ for the bi-cubic CY and $H=S_3$ for the triple tri-linear CY and arises due to the symmetries of the configuration matrices; it corresponds to permutations of the rows of the matrix $(B,C)$. The permutation groups $S_{r_B}$ and $S_{r_C}$ arise from permutations of the line bundles (columns) in $B$ and $C$ which of course do not affect the resulting monad bundle $V$. For the purpose of applying GAs, we have kept this degeneracy, as the ordering required for its elimination would significantly complicate the algorithm. Indeed the very notion of ordering by genotype, which would have to be performed even after each mutation, is somewhat at odds with the principle of GAs although 
it could in principle be included. However, redundancy in the models will be ultimately be eliminated from the list of perfect models found by the algorithm.\\[2mm]
For the purpose of implementing the GA we require an intrinsic value function $v(B,C)$ on the environment which measures any deviation from the required properties and, in the context of GAs, forms the basis of the fitness of a given individual. The various contributions to this value function are listed in Table~\ref{tab:intvaluemonad}.
\begin{table}[!h]
\begin{center}
\begin{tabular}{|l|l|l|}\hline
property&term in $v(B,C)$&comment\\\hline\hline
\varstr{19pt}{12pt}index match&$\displaystyle-\frac{2|{\rm ind}(V)+3|\Gamma|}{h M^3}$&${\rm ind}(V)$ computed from Eq.~\eqref{eq:monchern}\\\hline
\varstr{14pt}{8pt}anomaly&$\sum_{i=1}^h {\rm min}\left(c_{2i}(TX)-c_{2i}(V),0\right)$&$c_{2i}(V)$ computed from Eq.~\eqref{eq:monchern}\\\hline
\varstr{12pt}{7pt} bundleness&$-(d_{\rm deg}+1)$&$d_{\rm deg}$ from Eq.~\eqref{eq:deglocus}\\\hline
\varstr{12pt}{7pt} split bundle&$-n_{\rm split}$&$n_{\rm splits}$ as in Eq.~\eqref{eq:nsplits}\\\hline
\varstr{14pt}{7pt} equivariance&$\displaystyle -\sum_{L\subset B,C}{\rm mod}({\rm ind}(L^{\oplus m}),|\Gamma|)$&$L$ and $m$ is in Eq.~\eqref{eq:equivBC}\\\hline
\varstr{12pt}{7pt}  trivial bundle&$-n_{\rm trivial}$&$n_{\rm trivial}$ as in Eq.~\eqref{eq:nOX}\\\hline
\varstr{21pt}{11pt} stability $V$&$\displaystyle- \frac{{\rm max}(0,h^0(X,B)-h^0(X,C))}{hM^3}$&test from Eq.~\eqref{eq:Hoppetest}\\\hline
\varstr{21pt}{11pt} stability $V^*$&$\displaystyle - \frac{{\rm max}(0,h^0(X,B^*)-h^0(X,C^*))}{hM^3}$&test from Eq.~\eqref{eq:Hoppetest}\\\hline
\end{tabular}
\caption{\sf Contributions to the intrinsic value for the monad environment. The intrinsic value $v(B,C)$ is the sum of all eight terms.  The normalisation factors involve $M={\rm max}(b_{\rm max},c_{\rm max})$ and $h=h^{1,1}(X)$.}\label{tab:intvaluemonad}
\end{center}
\end{table}
A viable or perfect model $(B,C)$ is defined to be a model with intrinsic value $v(B,C)=0$. Given the definition of the value function in Table~\ref{tab:intvaluemonad}, such a state allows for a supersymmetric, anomaly-free completion, it is based on an $SO(10)$ GUT symmetry, has the correct chiral asymmetry of families and it passes the checks for bundle equivariance and poly-stability discussed earlier. Of course such perfect models might then still fail at a fairly elementary level, for example due to the presence of undesirable $\overline{\mathbf{16}}$-multiplets, a failure of stability or a failure of equivariance. Checking these properties requires cohomology calculations which, with the available methods in commutative algebra, are too time-consuming to be carried out during the GA evolution. However, experience from ``by-hand" model building~\cite{Anderson:2011ns,Anderson:2012yf,Anderson:2013xka} and from RL~\cite{Constantin:2021for} shows that a significant fraction of the perfect models also satisfy the more stringent cohomological constraints. In other words, while some of the conditions in Table~\ref{tab:intvaluemonad} are only necessary, they are still sufficiently strong~\footnote{It is conceivable that analytical formulae for monad cohomology can be found, in analogy with the analytical formulae known for line bundle cohomology~\cite{Constantin:2018hvl,Brodie:2019dfx,Brodie:2019pnz,Brodie:2019ozt}. Such formulae would facilitate setting up a more advanced environment which checks the full spectrum and a sufficient condition for stability.}.\\[2mm]
The monad environment we have just described was also used in the context of RL~\cite{Constantin:2021for}, where the value function in Table~\ref{tab:intvaluemonad} was used to determine the reward. It was shown in Ref.~\cite{Constantin:2021for} that RL is extremely successful on this environment, with trained policy networks efficiently leading to perfect models for 100\% of episodes, and many, previously unknown, models being discovered. Using the same environment allows a direct comparison between RL and GAs, which we carry out  in the next section.
In summary then, our environment is the same as that used for RL in Ref.~\cite{Constantin:2021for}, which has been realised as a MATHEMATICA~\cite{Mathematica} package.

\subsection{Genetic algorithm}
To realise the GA we choose the range of entries such that $b_{\rm max}-b_{\rm min}=2^{n_B}-1$ and 
$c_{\rm max}-c_{\rm min}=2^{n_C}-1$, representing every integer in $B$ and $C$ in a binary encoding, with $n_B$ and, respectively, $n_C$ bits, resulting in a bit sequence of total length $\ell=h(r_Bn_B+(r_C-1)n_C)$. The typical size of a population consists of $N_{\rm pop}\simeq 100 - 300$ states, each represented by such a binary encoding of the matrix $(B,C)$. The population is then evolved by crossing and mutation, as discussed in Section~\ref{sec:GA}. \\[2mm]
With the fitness function specified by the intrinsic value $v(B,C)$, we found that rank-weighting selection, elitism and a simple one-point cross breeding work well. 
The GA was implemented in two ways, as a MATHEMATICA~\cite{Mathematica} and as a Python package, in both cases being coupled to the MATHEMATICA package realising the environment. The two realisations lead to similar results.

\section{Monad bundles on the bicubic manifold}

\subsection{A typical GA run on the bicubic}

\begin{figure}[h]
     \centering
     \subfloat[][\centering Fitness histogram: number of individuals as a function of generation and fitness.]{\includegraphics[width=0.47\textwidth]{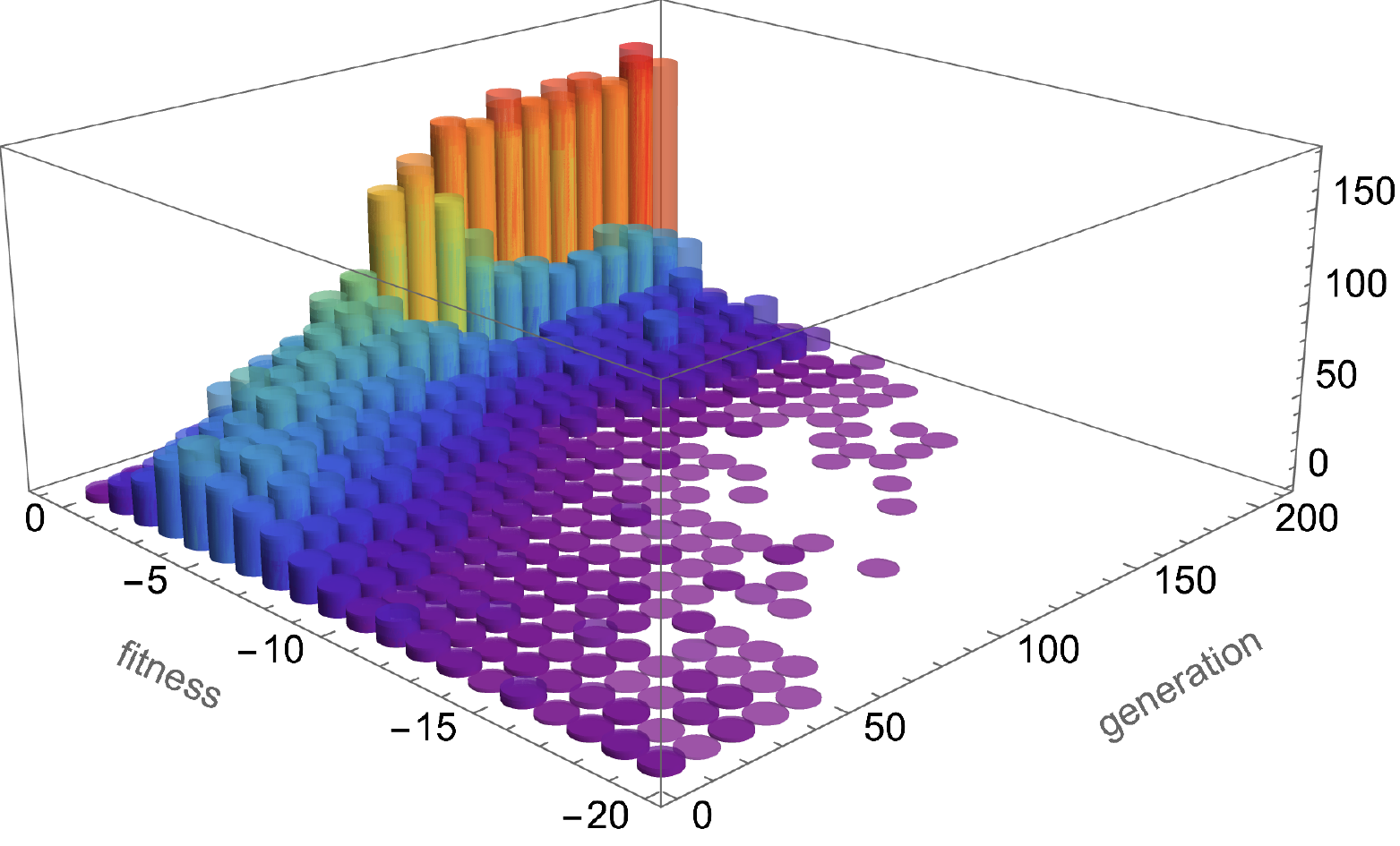}}
     \hspace{21pt}
     \subfloat[][\centering Fraction of perfect models vs generation.]{\includegraphics[width=0.47\textwidth]{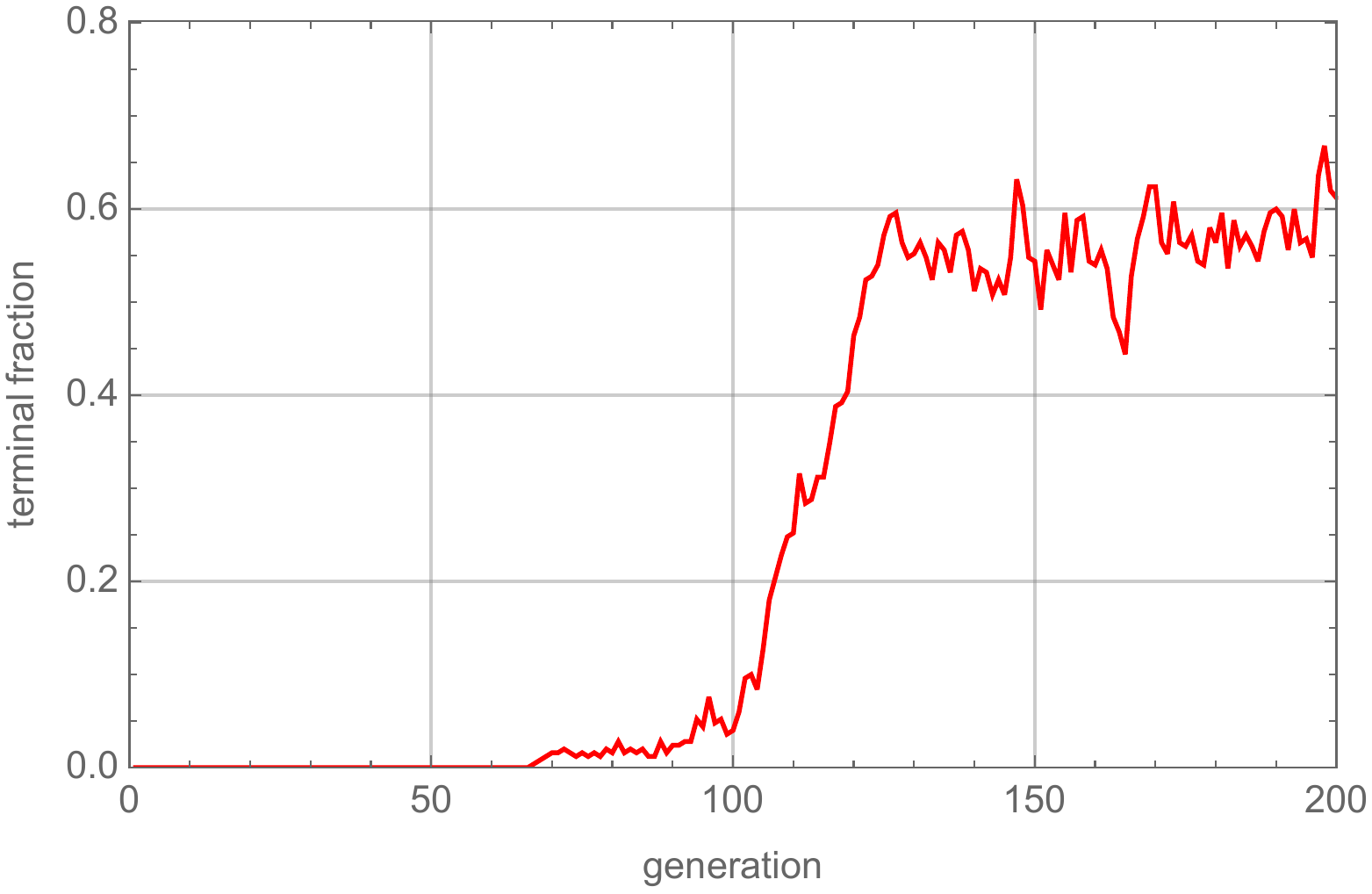}}
     \caption{Performance measures for a typical GA initialisation on the bicubic.}
     \label{figExampleBicubic}
\end{figure}

We first consider monad bundles on the bicubic manifold with ${\rm rk}(B)=6$ and ${\rm rk}(C) = 2$. We choose $b_{\rm min} = -3$, $c_{\rm min} = 0$ and $b_{\rm max}=4$, $c_{\rm max}=7$, such that each integer entry in the matrix $(B,C)$ can be represented by a $3$-bit binary sequence counting up from the fiducial points, $b_{\rm min}$ and $c_{\rm min}$. For instance, an entry of $-3$ in $B$ will be represented by the sequence $(0,0,0)$, while an entry of $4$ in $B$ corresponds to $(1,1,1)$. Similarly, an entry of $0$ in $C$ is represented as $(0,0,0)$, while an entry of $7$ in $C$ corresponds to $(1,1,1)$. \\[2mm]
Working with integers that can be encoded on 3 bits is convenient for the purpose of comparing the GA and RL techniques. However, it is possible and in fact easy to further increase the range of line bundle integers, as the corresponding augmentation in the computational time is not significant. \\[2mm]
As mentioned above, the environment has a large degeneracy. Equivalent models arise from permuting the two $\IP^2$-factors in the bicubic embedding, as well as from permuting the line bundles in $B$ and $C$. This amounts to a group of order $2!\cdot 6!\cdot2! = 2800$. Of course, this does not necessarily imply that every perfect model can be found in $2800$ different sites, since some permutations may leave particular matrices $(B,C)$ unchanged, but it nevertheless gives some idea of the degree of degeneracy present in the environment. \\[2mm]
It is useful to illustrate the performance of the GA graphically. We set the size of the population to $N_{\rm pop}=250$ randomly initialised states and let the GA run for $200$ generations with a mutation rate of $0.004$ and the parameter $\alpha$ from Eq.~\eqref{eq:Pk} set to $\alpha=3$. The time taken by such a run is relatively small ($<\!\!100$ seconds on a laptop). 
Figure~\ref{figExampleBicubic} shows a typical evolution of the population. The histogram on the left shows how the population progresses towards greater fitness (intrinsic value). The plot on the right shows the fraction of perfect models in the population, that is, the fraction of models with a value $v(B,C)=0$. By generation $150$ more than half of the population corresponds to perfect models. For this run, a total of $200\times 250$ states have been visited, many of them multiple times. Of these $12,665$  correspond to perfect models with $48$ being distinct. After eliminating redundancies, $18$ non-equivalent models remain. 

\begin{figure}[h]
     \centering
    {\includegraphics[width=0.45\textwidth]{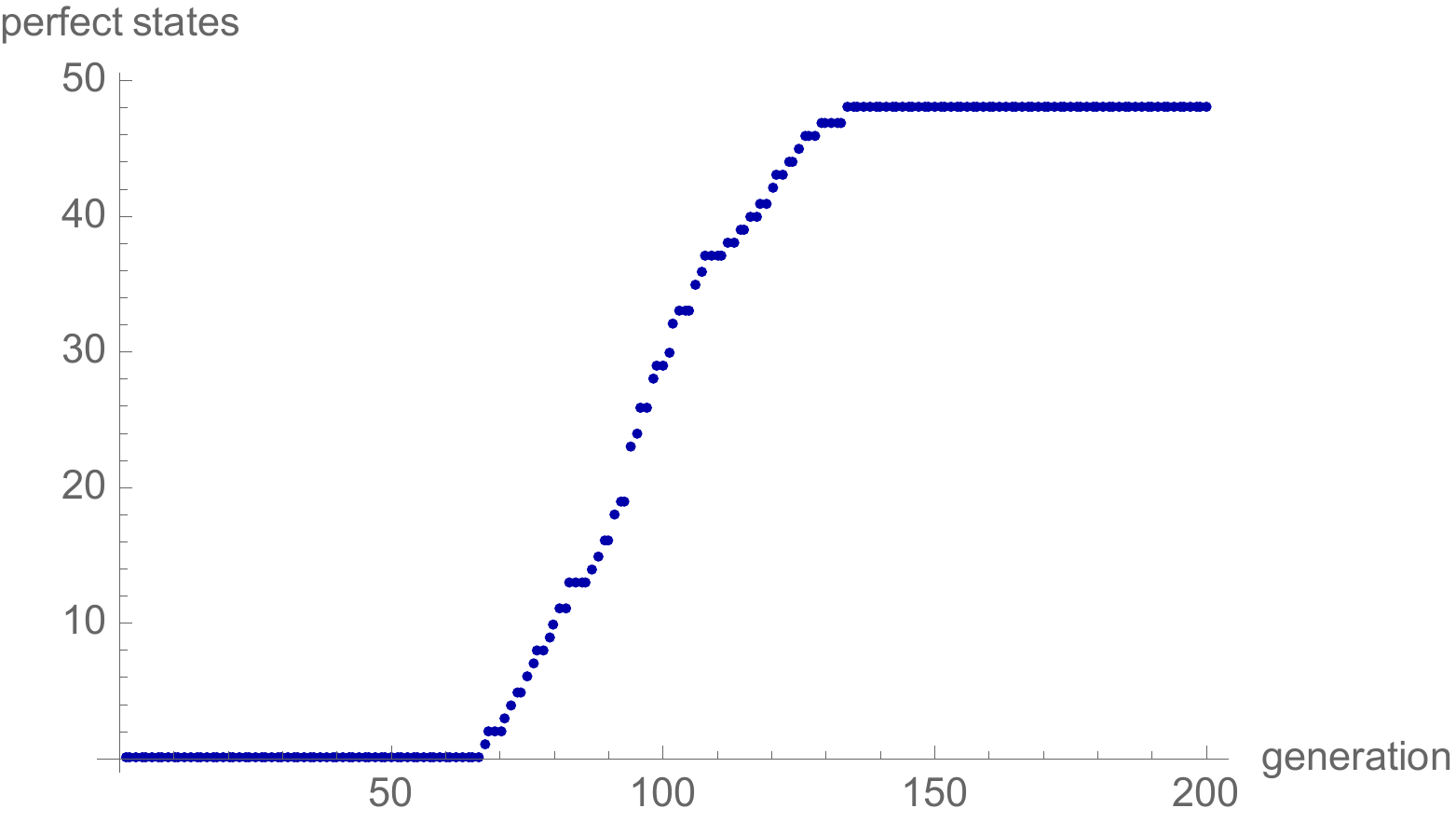}}
       \caption{Saturation of the number of perfect states found in a typical GA run on the bicubic.}
     \label{fig:BicubicPerfectStates}
\end{figure}

\noindent The fraction of perfect models can be easily increased by lowering the mutation rate. However, as one might anticipate from the discussion in the introduction, this also induces a degree of stagnation, with a tendency for the same states to emerge repeatedly.
\\[2mm]
A few remarks are in order. Firstly, to appreciate the performance of the algorithm, note that the size of the search space is 
\[
8^{14}\simeq 4.4\times 10^{12}\;.
\]
By comparison, the number of states visited in the above run, namely $50,000$,  represents only a tiny fraction of the space. However, the GA was capable of finding $48$ perfect states, while a random search over millions of states would typically lead to no perfect states at all. \\[2mm]
Secondly, as already noted, the GA has a tendency of visiting the same states multiple times. It is interesting to plot the total number of perfect states found after $n$ generations as a function of $n$. For our illustrative run such a plot is shown in Figure~\ref{fig:BicubicPerfectStates}, which suggests that there is no additional benefit in letting the population evolve beyond a certain generation ($n\simeq 150$). 

\noindent Finally, it is useful to compute the degeneracy of the $18$ states that remain after removing redundancies. By performing all the allowed permutations a number of $19,080$ states are obtained. This should be compared with the product $18\times 2800=50,400$ which turns out to be an overestimate by more than a factor of $2$. Moreover, what this computation indicates is that a single run of the GA is not enough if the aim was to find all the perfect states available in the environment and that $1,000$ further runs, which would take about 1 day on a single machine, would be just about enough to find the other redundant representations of the $18$ states found in the first run. Of course, many more new states, not related to the $18$ by permutations, would be expected to arise in such a search, which implies that a comprehensive search would require several, possibly tens of thousands of GA runs. With $10,000$ runs this would amount to exploring $\sim 0.01\%$ of the environment.

\subsection{A quasi-comprehensive GA search}
\begin{figure}[h]
     \centering
    {\includegraphics[width=0.59\textwidth]{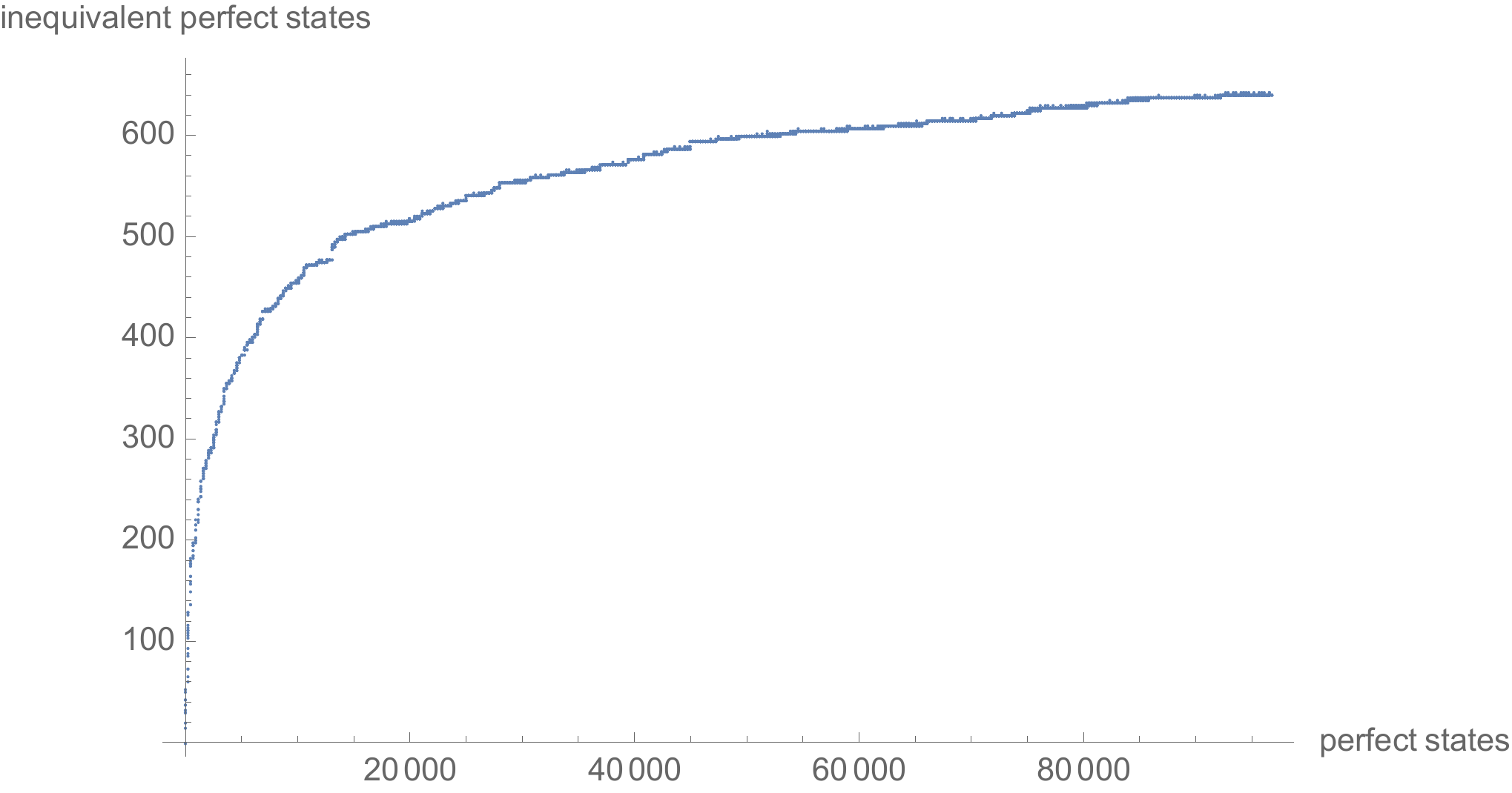}}
       \caption{GA scan results. Saturation of the redundancy reduced number of perfect states.}
     \label{fig:BicubicGASaturation}
\end{figure}

The illustrative results presented above show that GAs can be efficiently used to identify monad bundle models on the bicubic manifold that satisfy the phenomenological requirements discussed in Section~\ref{sec:heterotic_intro}. A natural question to ask is whether the GA can find {\it all} (or almost all) perfect models available within the environment. Answering this question is difficult, mainly because the environment is too large to be subjected to a systematic scan which would ensure that no perfect state was missed. The computational time needed for a systematic scan within the environment bounds can be easily estimated to about 200 core years, which although attainable, is by no means something that one would easily invest in. \\[2mm]
A partial answer to the question can be achieved by making use of the redundancy present in the environment. The procedure is as follows. We perform 10,000 runs keeping the same settings as above, namely $N_{\rm pop}=250$, the length of a run equal to $200$ generations, a mutation rate of $0.004$ and $\alpha=3$. The required computation time for this is about $10$ core days\footnote{The scan was carried out on the Hydra Computer Cluster in the Dept.~of Physics at the University of Oxford.}. We extract the number of perfect states found after $n$ runs and then ask how many of these are left after removing redundancies. Now, if the GA is capable of finding almost all the states, then the expectation would be that the redundancy reduced number of perfect states after $n$ runs would saturate as a function of the total number of perfect states found after $n$ runs. In other words the GA is only finding versions of models that it has found before.  The plot in Figure~\ref{fig:BicubicGASaturation} confirms this expectation. \\[2mm]

\begin{figure}[h]
     \centering
      \subfloat[][\centering Number of inequivalent perfect states found as a function of the number of states visited.]{\includegraphics[width=0.47\textwidth]{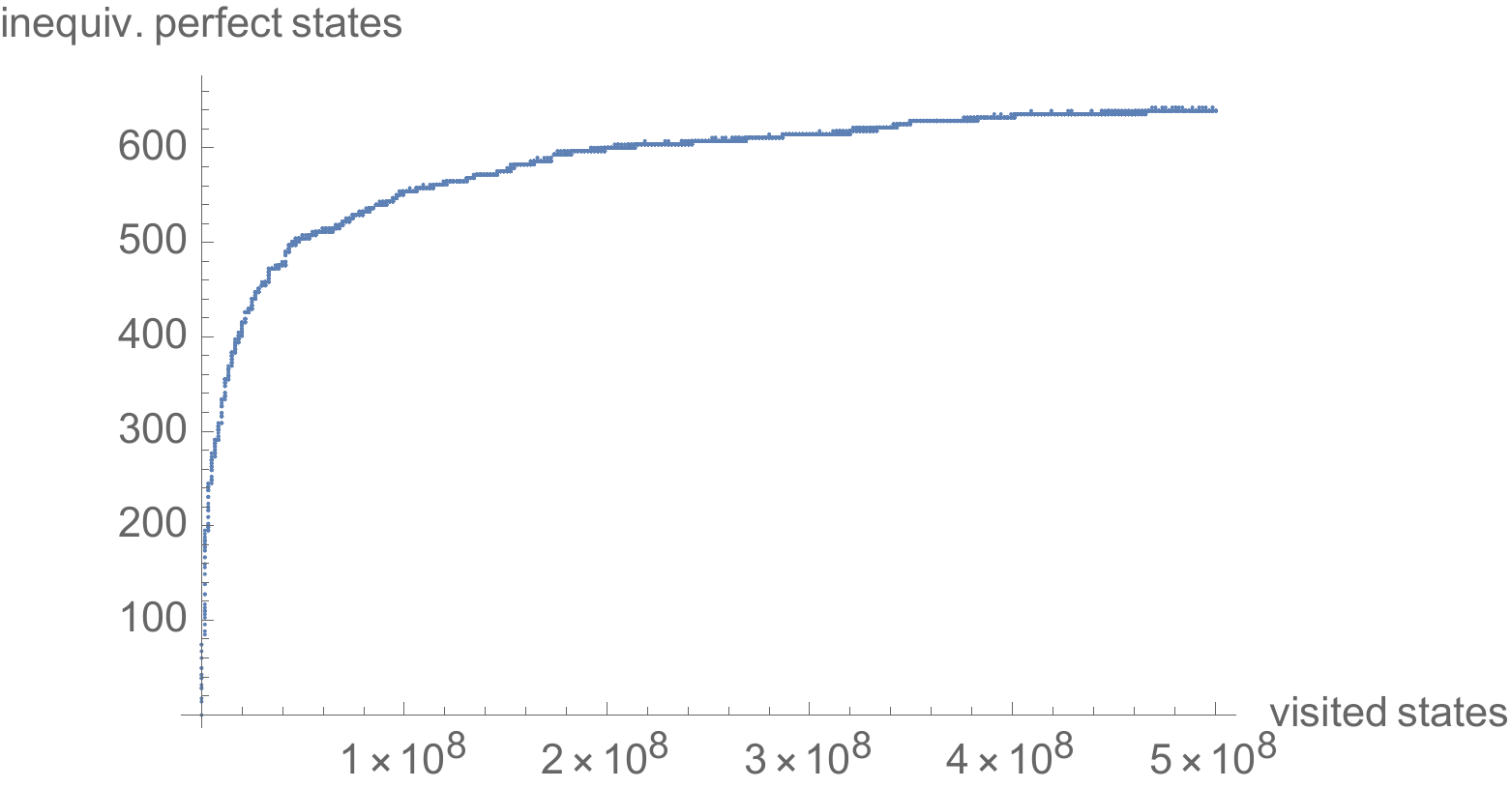}}
     \hspace{21pt}
     \subfloat[][\centering Number of perfect states found as a function of the number of states visited.]{\includegraphics[width=0.47\textwidth]{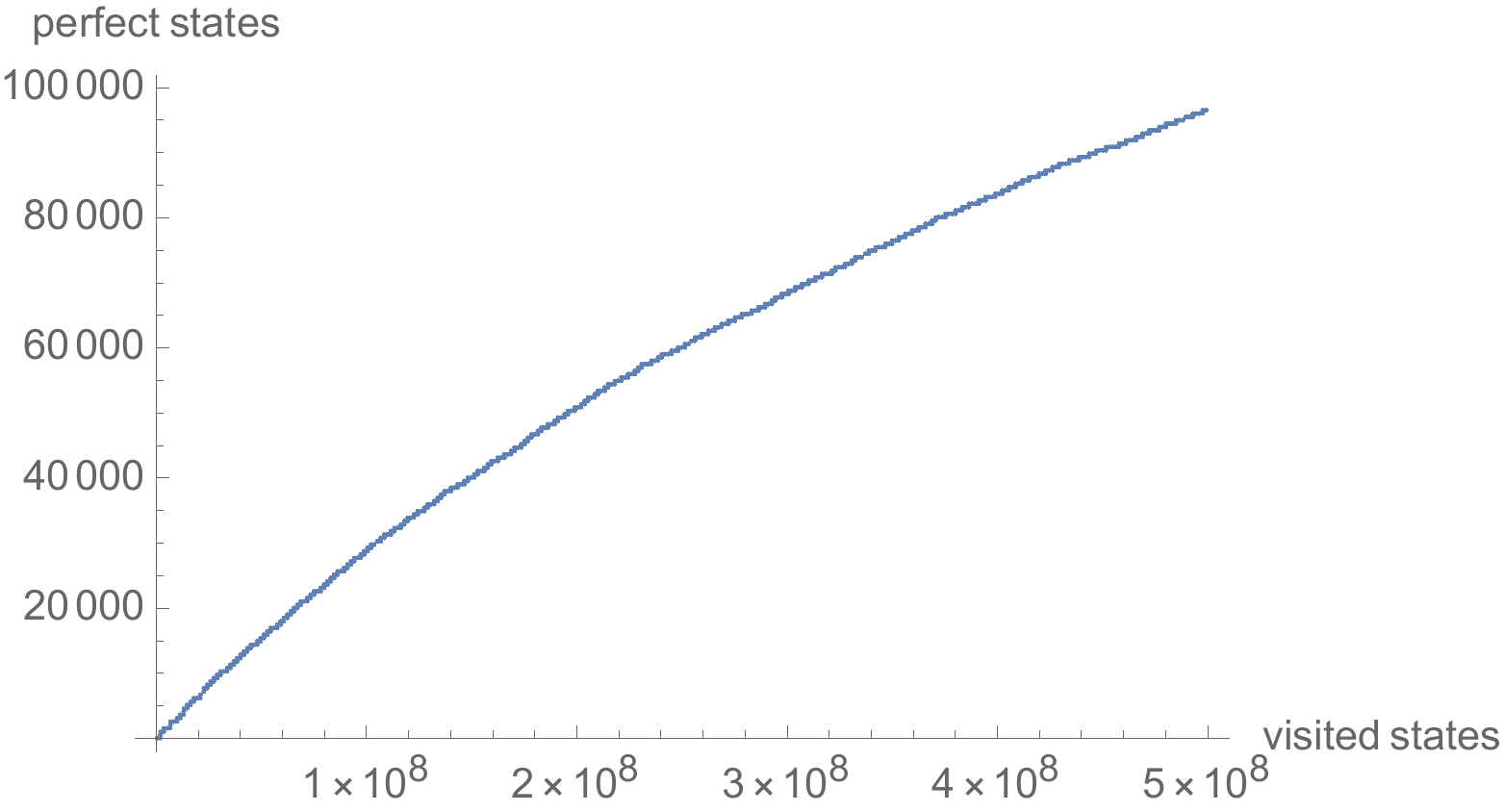}}
     \caption{GA search results on the bicubic. The number of inequivalent perfect states saturates, while the total number of perfect states found is still far from saturation after 10,000 GA runs.}
     \label{fig:BicubicGASaturation1_2}
\end{figure}

\noindent The search lead to a number of $96,705$ perfect states, out of which $639$ are inequivalent (not related by a permutation). Performing all possible permutations on the $639$ states leads to $481,680$ perfect states that must be present in the environment. If allowed more time, the GA would of course be expected to find all of these. However, the saturation plot in Figure~\ref{fig:BicubicGASaturation} as well as the plots in Figure~\ref{fig:BicubicGASaturation1_2} suggest that there are not many more inequivalent states to be found. We take this as an indication that the search has already successfully found most of the inequivalent perfect states. 

\subsection{Enlarging the search space}
\begin{figure}[h]
     \centering
     {\includegraphics[width=0.4\textwidth]{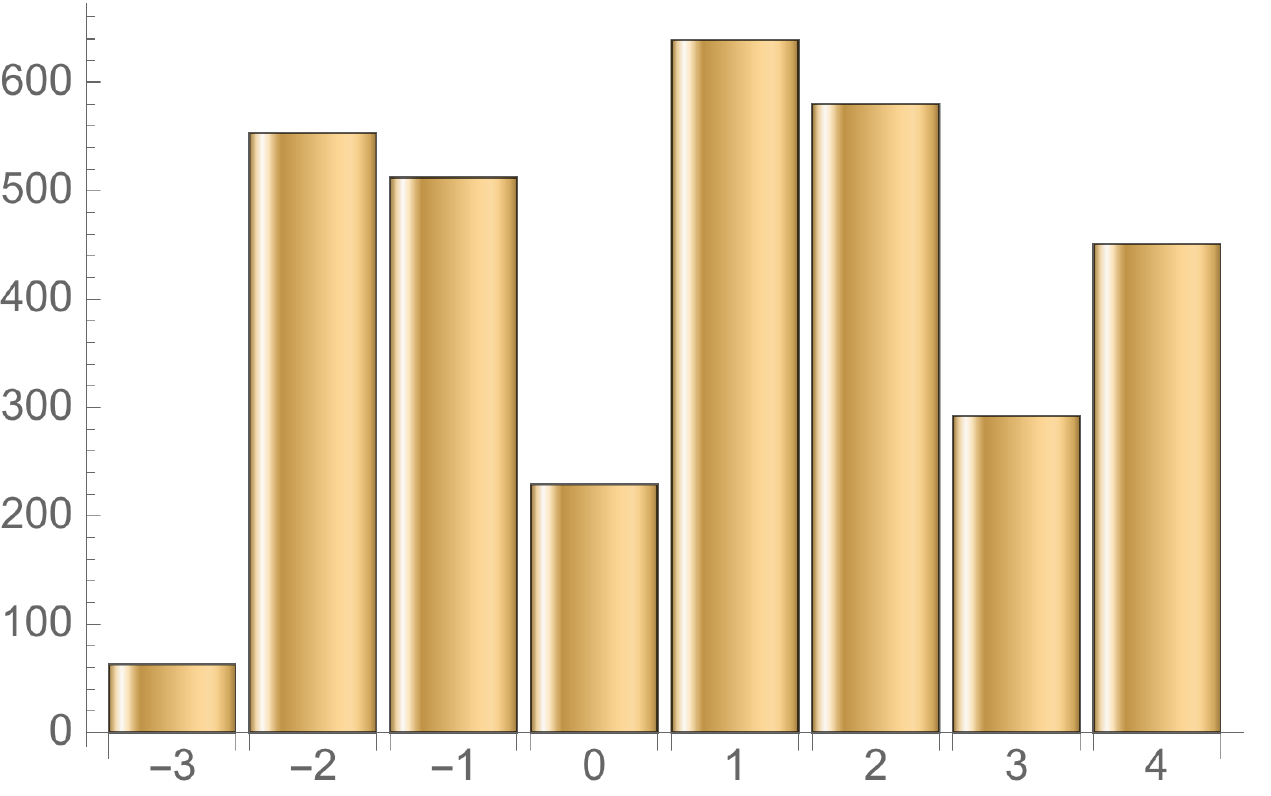}}
     \hspace{21pt}
     {\includegraphics[width=0.4\textwidth]{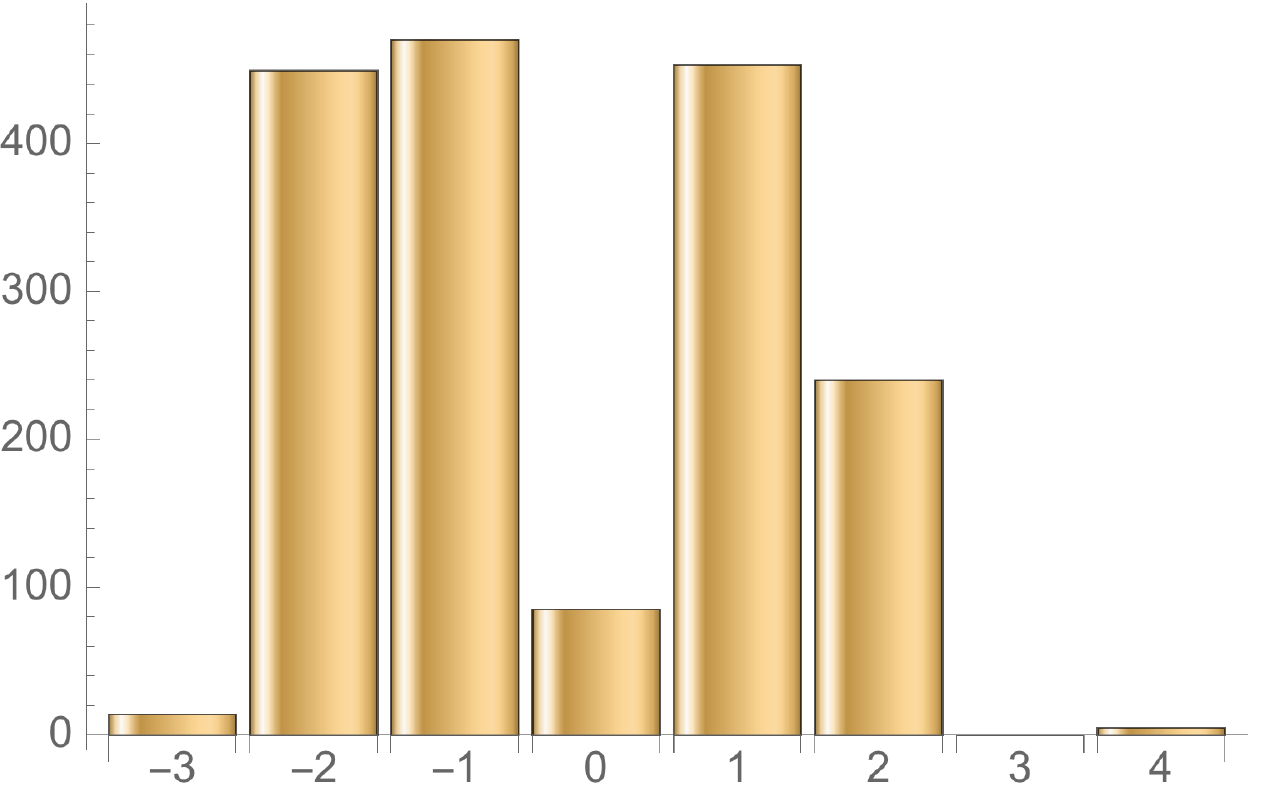}}
     \caption{Bar charts showing the number of inequivalent perfect states $(B,C)$ that have a certain integer present in $B$ (left chart) and in $C$ (right chart).}
     \label{fig:BarCharts}
\end{figure}

So far the minimal and the maximal values of the entries in $B$ and $C$ were fixed such that all integers can be encoded on 3 bits. Upgrading to integers that can be encoded on 4 bits leaves the computational time required for a GA run virtually unchanged. Thus enlarging the range of line bundle integers in $B$ and $C$ would not pose a problem for the performance of the GA implementation used here. But is there a good reason for enlarging the environment? \\[2mm]
To answer this question we need to see if the perfect models found are well within the boundary of the environment or if a significant fraction of them lies close to the boundary. The bar charts in Figure~\ref{fig:BarCharts} show that in fact most of the models lie on the boundary (for instance, $451$ out of the $639$ inequivalent models include the integer $4$ in $B$). This suggests the need for increasing further the entries in $B$ and $C$. 

\noindent However, before embarking on such an enterprise, it is worth looking deeper into the quality of the monad bundles  found. The requirement of stability imposes that $h^0(V)=h^3(V)=0$. Moreover, in order to avoid the presence of $\overline{\mathbf{16}}$-multiplets, the bundle has to satisfy $h^2(V)=0$. These cohomological constraints are  too computationally intensive to have included them in the search, however they can be carried out on the set of $639$ inequivalent models found on the bicubic manifold.  \\[2mm]
\begin{figure}[h]
     \centering
    {\includegraphics[width=0.4\textwidth]{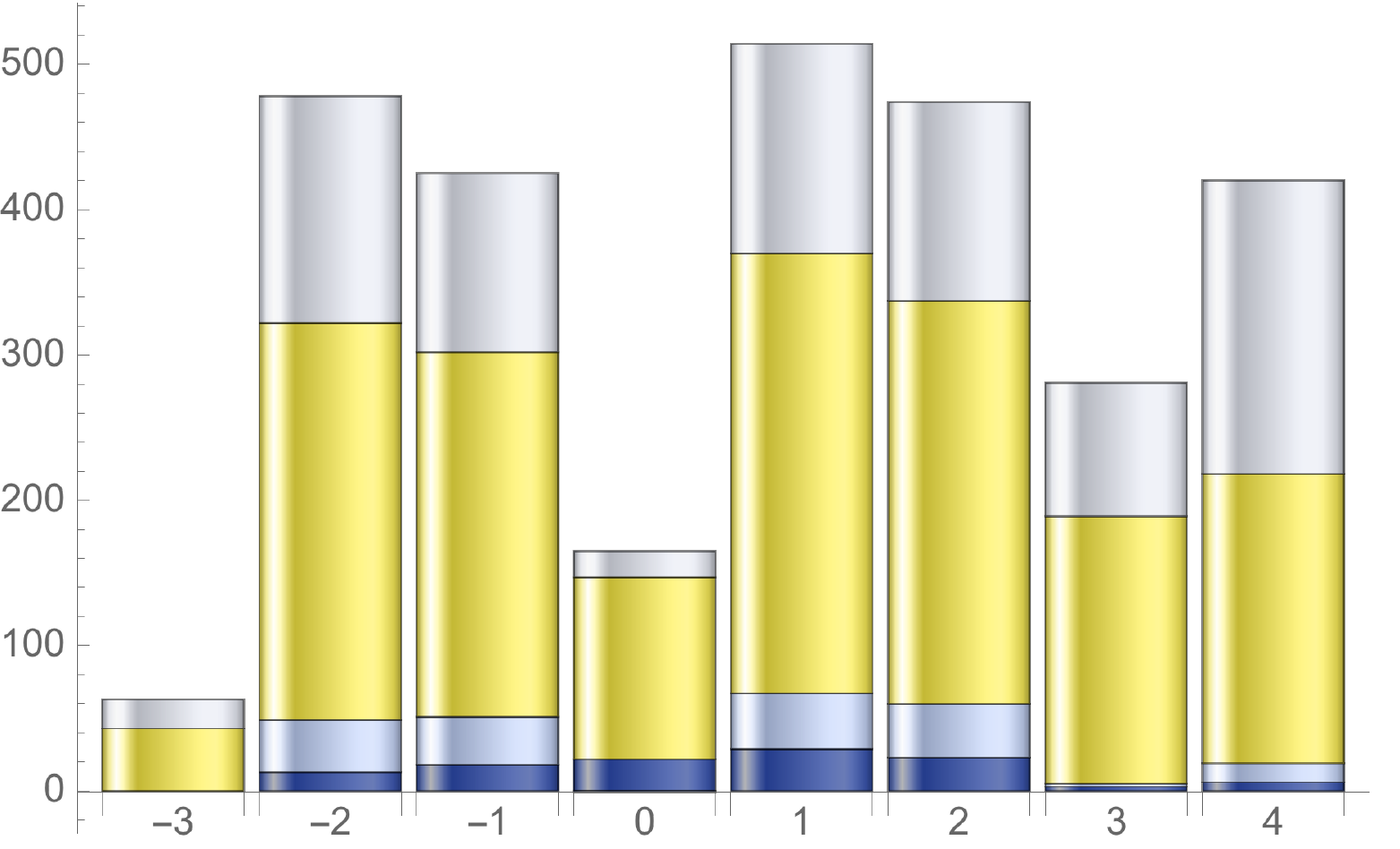}}
    \hspace{21pt}
        {\includegraphics[width=0.4\textwidth]{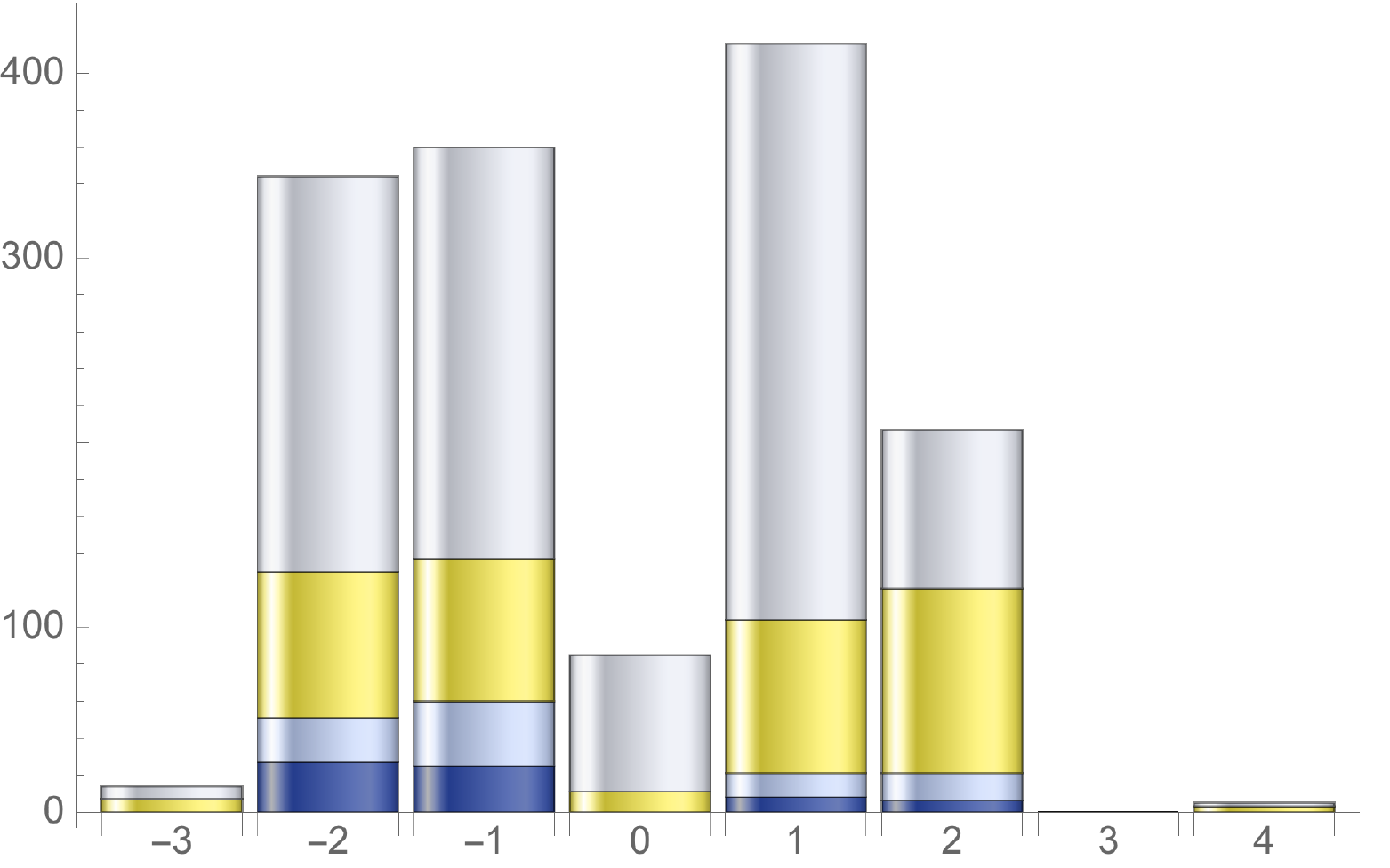}}
      \vspace{4pt}
              {\includegraphics[width=0.4\textwidth]{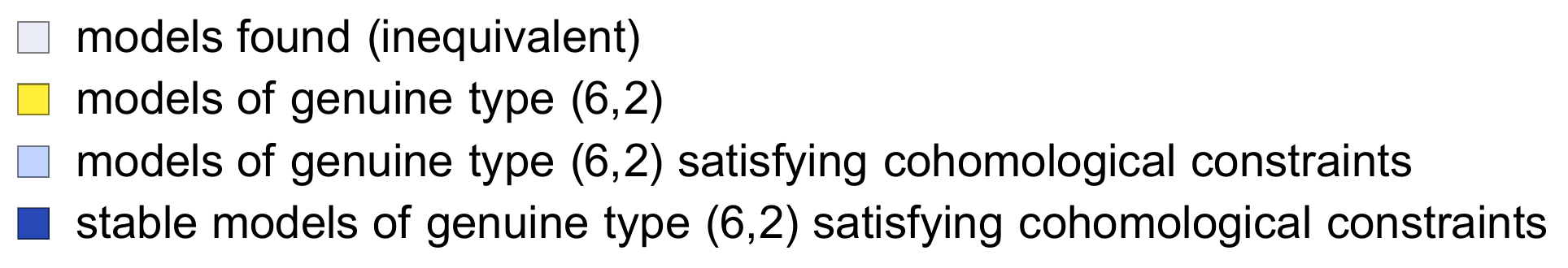}}
       \caption{Distribution of models across the range of integers in $B$ (top chart) and $C$ (bottom chart). The number of models is significantly reduced when additional requirements are imposed. Very few models are left on the boundary of the environment.}
     \label{fig:BarChartAfterChecks}
\end{figure}

\noindent Moreover, many of the models found contain a repeated line bundle in $B$ and $C$. While this is not a problem in itself, such models are equivalent to monad bundles with ${\rm rk}(B)=5$ and ${\rm rk}(C)=1$. We eliminate these from the discussion, which leaves $214$ models of genuine ${\rm rk}(B)=6$ and ${\rm rk}(C)=2$ type. Out of these, $67$ have the right cohomologial properties mentioned above. Further stability checks reduce this number to $29$ models.
The distribution of entries in $B$ for the remaining $29$ models is shown in blue in Figure~\ref{fig:BarChartAfterChecks} and indicates that most of the models have all the line bundle entries in the range $\{-2,-1,0,1,2\}$. In view of this, the expectation is that very few, if any, additional models satisfying the additional constraints would be found by further enlarging the size of the environment. 

\subsection{GA scan versus RL scan}
\begin{figure}[h]
     \centering
    {\includegraphics[width=0.4\textwidth]{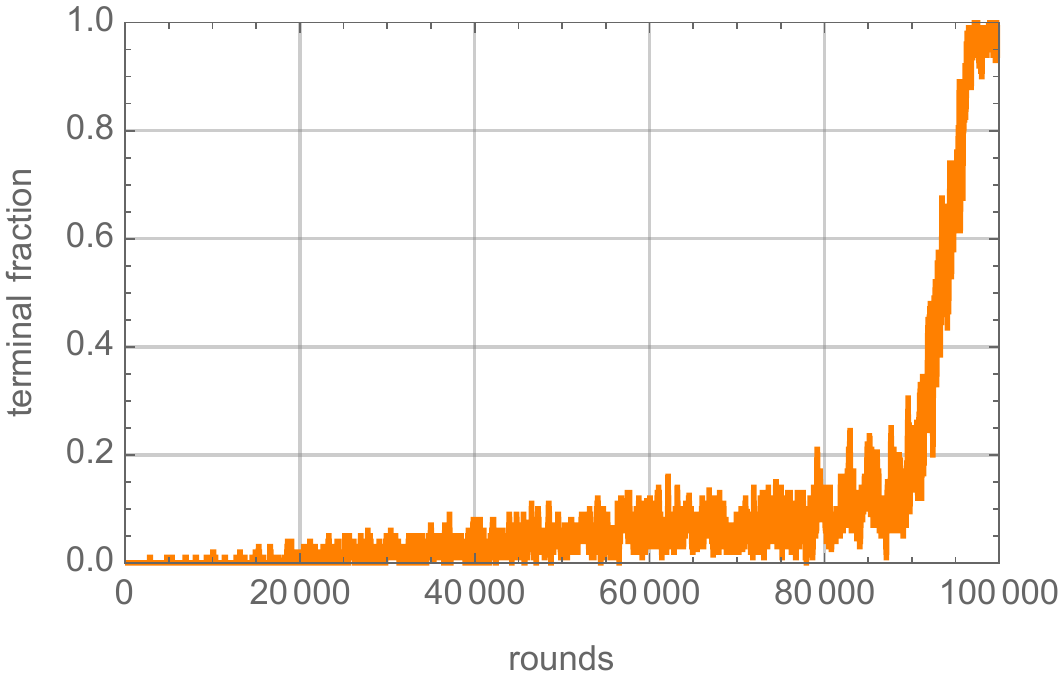}}
    \hspace{21pt}
        {\includegraphics[width=0.4\textwidth]{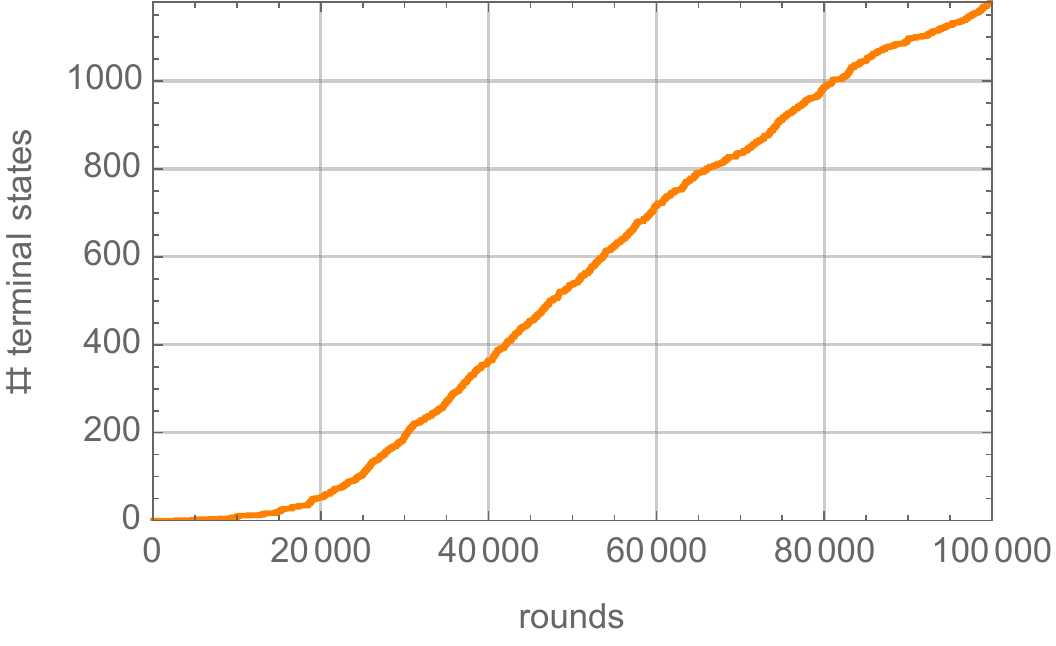}}
       \caption{RL training metrics on the bicubic manifold. After $100,000$ rounds every episode ends up in a terminal state.}
     \label{fig:RLTraining}
\end{figure}

In Ref.~\cite{Constantin:2021for} a similar search for physically viable models on the bicubic manifold with monad bundles was performed using methods of reinforcement learning (RL). The same fitness function $v(B,C)$ as described above was used in that context to compute the intrinsic value of states $(B,C)$. When moving from one state to another, the RL agent receives a reward equal to the difference between the intrinsic values of the two states if the difference was positive or a fixed penalty if the difference was negative. The search is divided into episodes of fixed maximal length and along each such episode the return $G_t$ of a state $s_t$ is computed as
\begin{equation}\label{returndef}
 G_t=\sum_{k\geq 0} \gamma^k r_{t+k}\; ,
\end{equation} 
where $r_{t}$ is the reward (penalty) associated with the step $s_t\rightarrow s_{t+1}$ and the discount factor $\gamma$ is a subunitary number close to $1$. The aim of the game is then to devise a strategy of finding a path that would maximise the return of any randomly chosen starting point, by eventually landing on a perfect state (also called a terminal state in the RL  context). \\[2mm]
More details can be found in Ref.~\cite{Constantin:2021for}. What is important for the present discussion is that, although the RL and GA methods rely on the same intrinsic value function, their underlying philosophies are very different, which makes for an interesting comparison. \\[2mm]
For the RL we use a policy network trained on episodes of maximal length $64$. After $100,000$ rounds of training, a phase which takes several hours on a laptop, the RL agent is capable of finding terminal states from virtually any random starting point within episodes of average length $\sim\!\!30$. The training curves are shown in Figure~\ref{fig:RLTraining}.

\noindent In order to get a preliminary assessment of the performance of the trained network we let it run for $10,000$ episodes which takes about $16$ minutes on a single machine. The run produced $2460$ terminal states, out of which only $72$ are distinct. The repetitions are not evenly distributed over the $72$ distinct states. In fact, one of these states accounts for almost half of the terminal states found. Incidentally, this particular state corresponds to a monad bundle that can be reduced to a ${\rm rk}(B)=5$ and ${\rm rk}(C)=1$ case. The genuine ${\rm rk}(B)=6$ and ${\rm rk}(C)=2$ are found with much smaller frequencies, most of them appearing only once in the dataset. Thus, although the network is capable of finding terminal states, a small number of these sates have very large basins of attraction and are found much more often than the others. This is a source of inefficiency for the RL implementation. 
\begin{figure}[h]
     \centering
    {\includegraphics[width=0.59\textwidth]{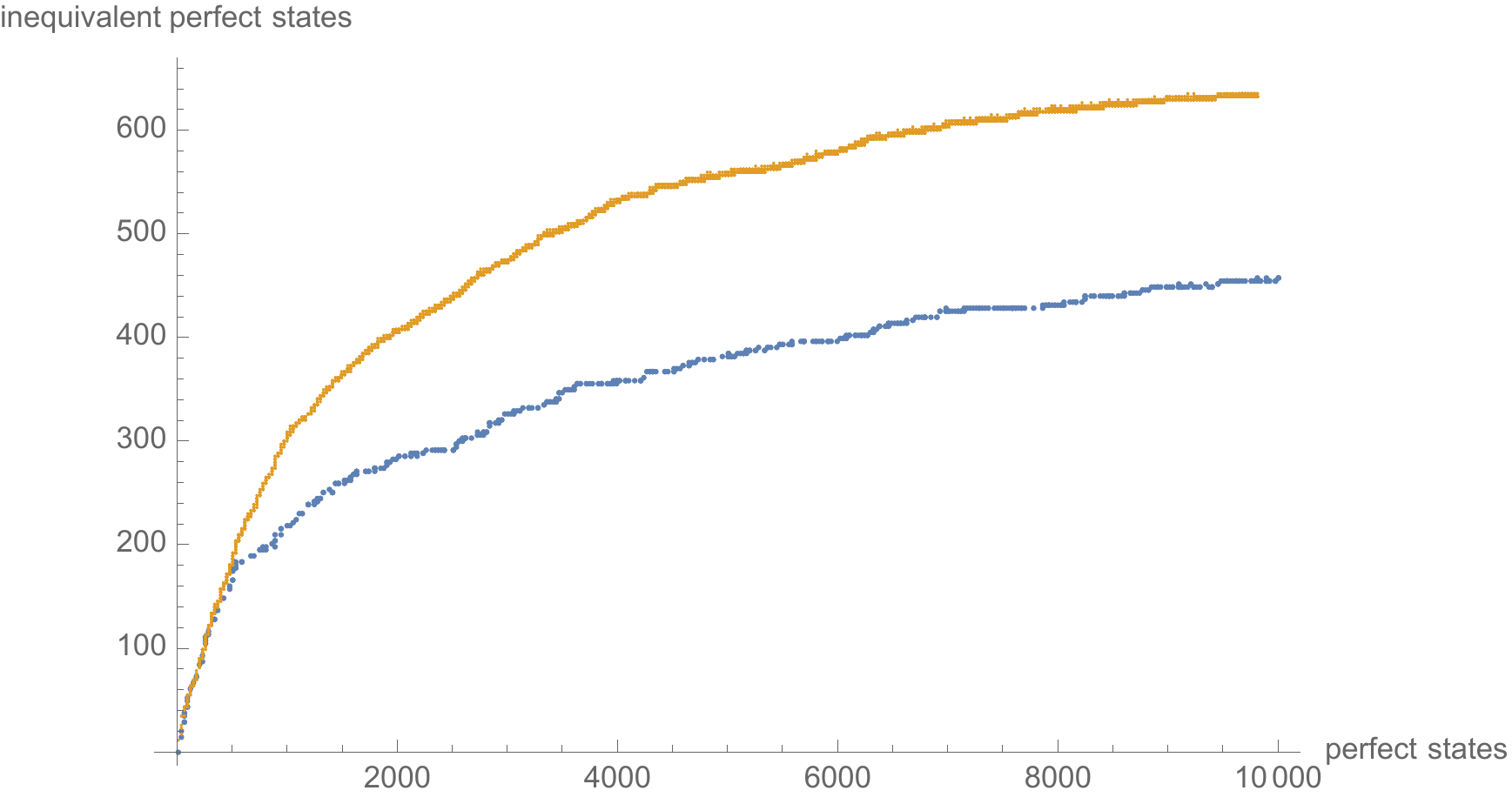}}
       \caption{Saturation of the redundancy reduced number of perfect states. Orange curve: RL search. Blue curve: GA search. The RL search took $35$ core days, while the part of the GA search shown here took only $1$ core day.}
     \label{fig:BicubicRLSaturation}
\end{figure}

\begin{figure}[h]
     \centering
      \subfloat[][\centering Number of inequivalent perfect states found as a function of the number of states visited.]{\includegraphics[width=0.47\textwidth]{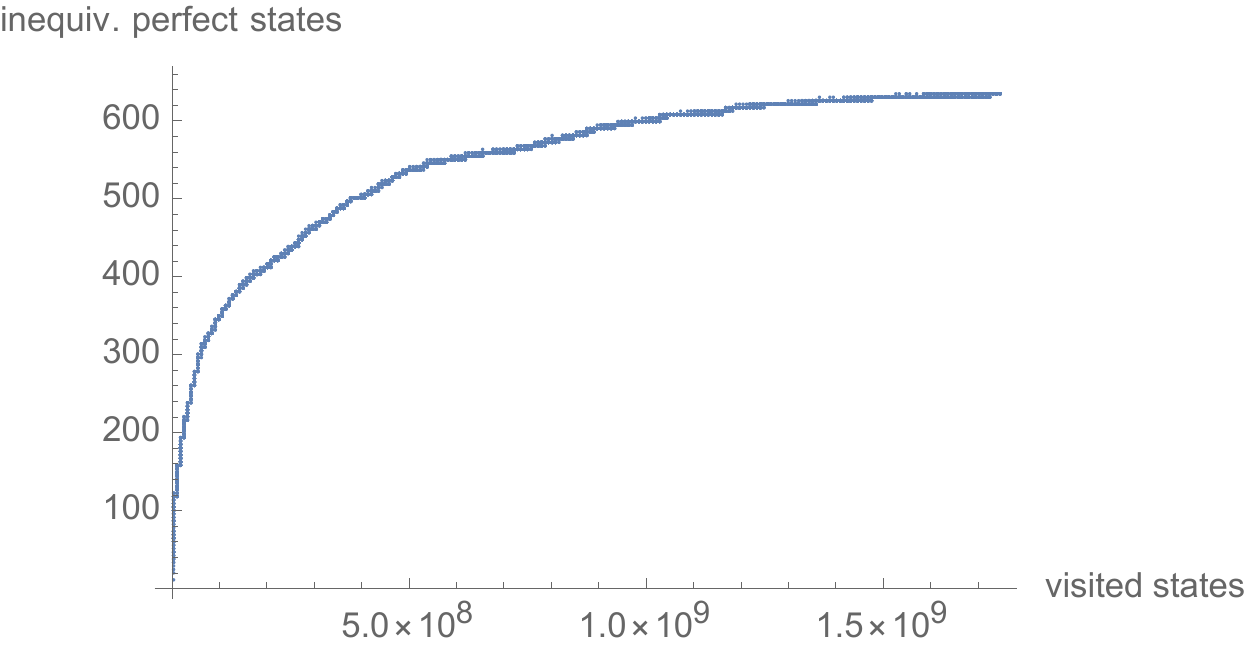}}
     \hspace{21pt}
     \subfloat[][\centering Number of perfect states found as a function of the number of states visited.]{\includegraphics[width=0.47\textwidth]{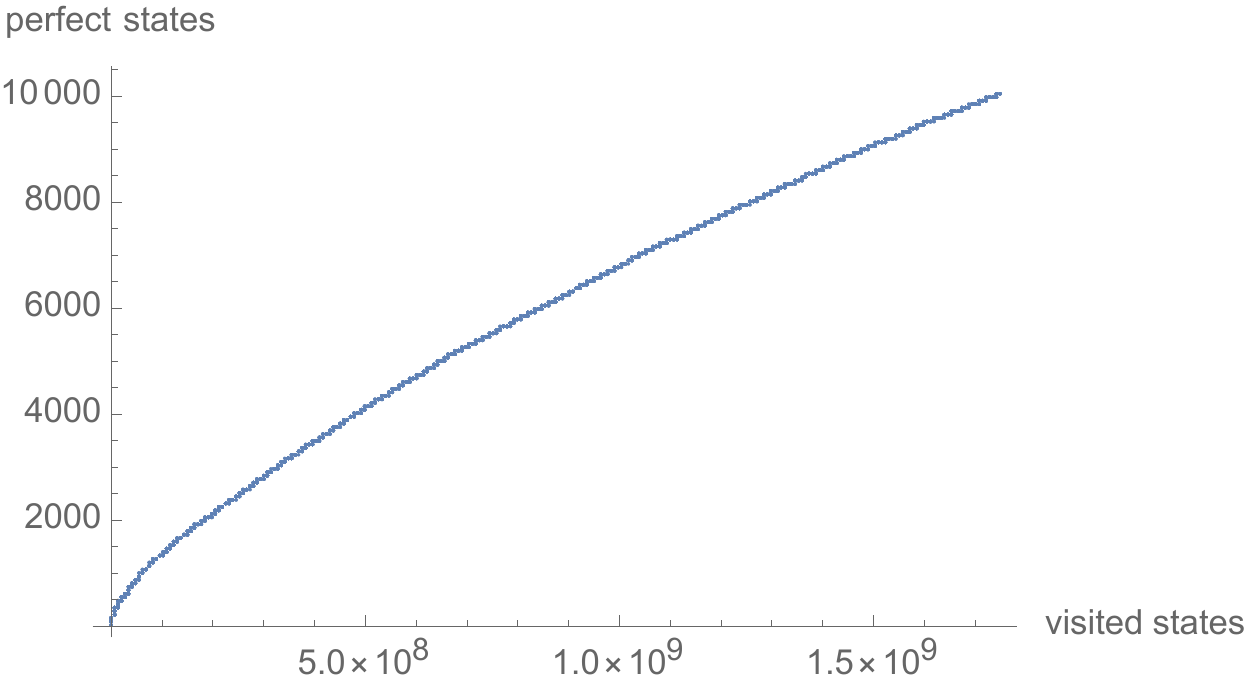}}
     \caption{RL search results on the bicubic. The number of inequivalent perfect states saturates, while the total number of perfect states found is still far from saturation after $1.7\cdot 10^9$ RL episodes.}
     \label{fig:BicubicRLSaturation1_2}
\end{figure}

\noindent We let the trained network search for terminal states from $1.7\cdot 10^9$ random initial states. This search, which took $35$ core days produced $\sim\!\!10,000$ perfect states, out of which $643$ were inequivalent. By comparison, the computational time for the GA scan was $10$ core days and lead to $\sim\!\!100,000$ perfect states, out of which $639$ were inequivalent.  The saturation curve of the number of inequivalent models as a function of the total number of models found at various stages of the RL search is shown in Figure~\ref{fig:BicubicRLSaturation} together with the corresponding curve for the first $10,000$ perfect states found during the GA search. Figure~\ref{fig:BicubicRLSaturation1_2} shows the progression of the number of (inequivalent) models found with the number of states visited.

\noindent A number of remarks can now be made. Firstly, the GA search is more efficient at finding perfect states than the RL search. Roughly the same number of inequivalent perfect states have been found in the two searches, though the RL search took $3.5$ times longer than the GA. Secondly, the GA is more prolific, by more than an order of magnitude compared to the RL search, at finding redundant perfect states. This is very likely a by-product of crossover, which is prone to producing redundant states by permuting various parts of the genome. Note that various techniques such as crowding penalties might be considered to remove this inefficiency.
 
\noindent Finally, there is a substantial overlap between the inequivalent models found in the two searches, with only $\sim\!\!50$ models in each complement. This brings the total number of inequivalent models to $689$. The fact that two such different methods lead to virtually the same dataset of models is another even stronger indication that it has been possible to achieve a high degree of comprehensiveness by scanning only a tiny fraction of the environment.\\[2mm]
While both methods produce nearly identical data sets of perfect models after sufficient running time their trajectories during evolution/training appear to be quite different. Using nonlinear visualisation techniques such as Sammon maps we find that the set of perfect models found at an early stage is quite different for the two methods. In particular, as illustrated by Fig.~\ref{fig:BicubicRLSaturation}, GAs tend to produce a significantly larger degree of redundancy than RL.

\section{Monad bundles on the triple trilinear manifold}

\subsection{A typical GA run on the triple trilinear}

As in the previous section we consider monad bundles with ${\rm rk}(B)=6$ and ${\rm rk}(C) = 2$ and set $b_{\rm min} = -3$, $c_{\rm min} = 0$ and $b_{\rm max}=4$, $c_{\rm max}=7$, such that each integer entry in the matrix $(B,C)$ can be represented by a $3$-bit binary sequence. 
For the triple trilinear manifold the degeneracy of the environment is even larger than in the case of the bicubic, due to the six possible permutations of the three $\IP^2$-factors in the ambient space. The corresponding group has order $8640$.\\[2mm]
As in the previous case, we illustrate the performance of the GA graphically. We keep the size of the population to $N_{\rm pop}=250$ randomly initialised states. The GA needs to run a little longer until a significant fraction of the population corresponds to perfect states. Evolutionary episodes of $400$ generations with a mutation rate of $0.004$ and $\alpha=4$ seem to work well. The time taken by such a run is relatively small ($\sim\!\!450$ seconds on a single machine). 

\begin{figure}[h]
     \centering
     \subfloat[][\centering Fitness histogram. Height: number of individuals. Front axis: fitness (intrinsic value). Depth: generation number.]{\includegraphics[width=0.47\textwidth]{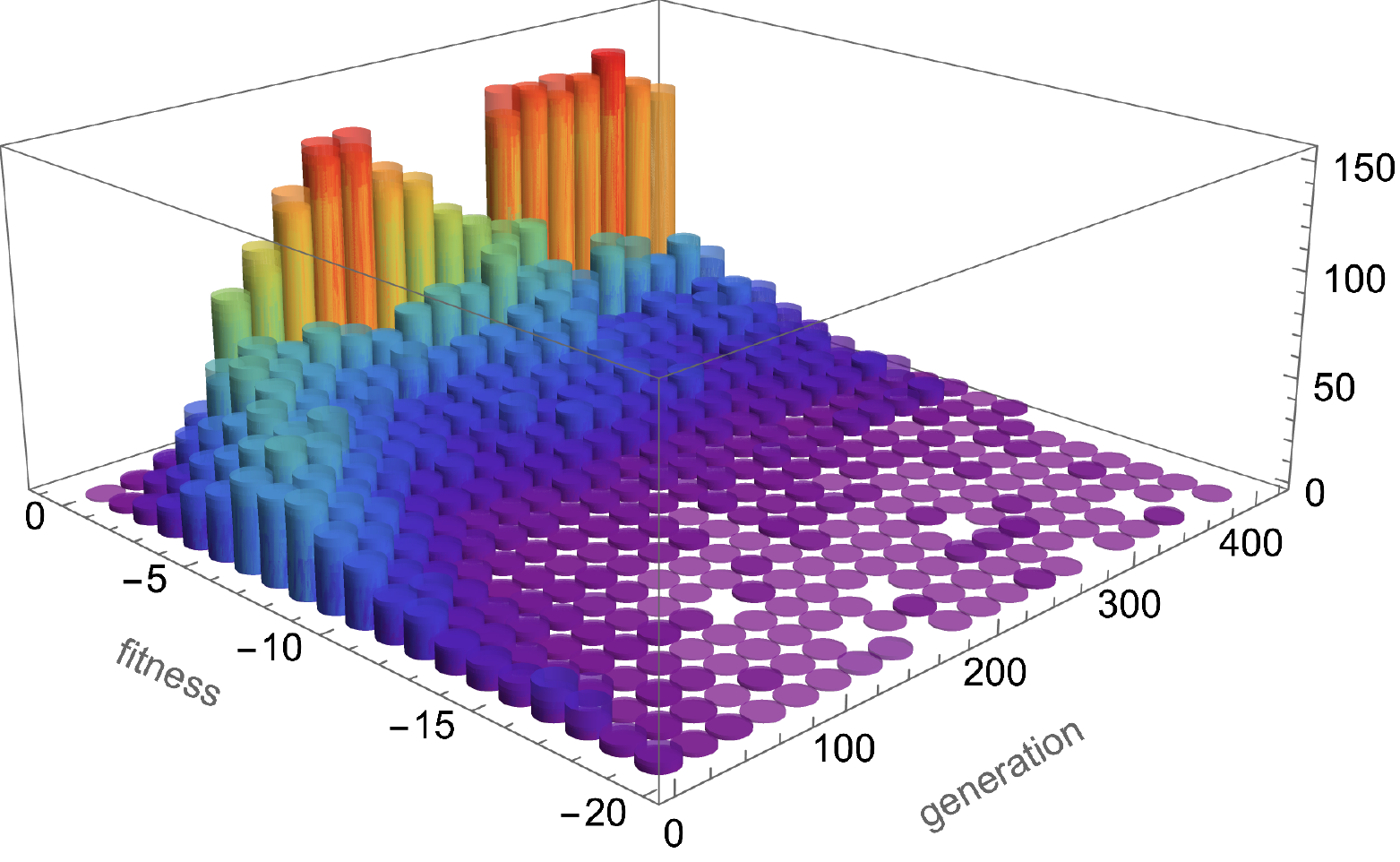}}
     \hspace{21pt}
     \subfloat[][\centering Fraction of models that achieved `perfection' vs generation.]{\includegraphics[width=0.47\textwidth]{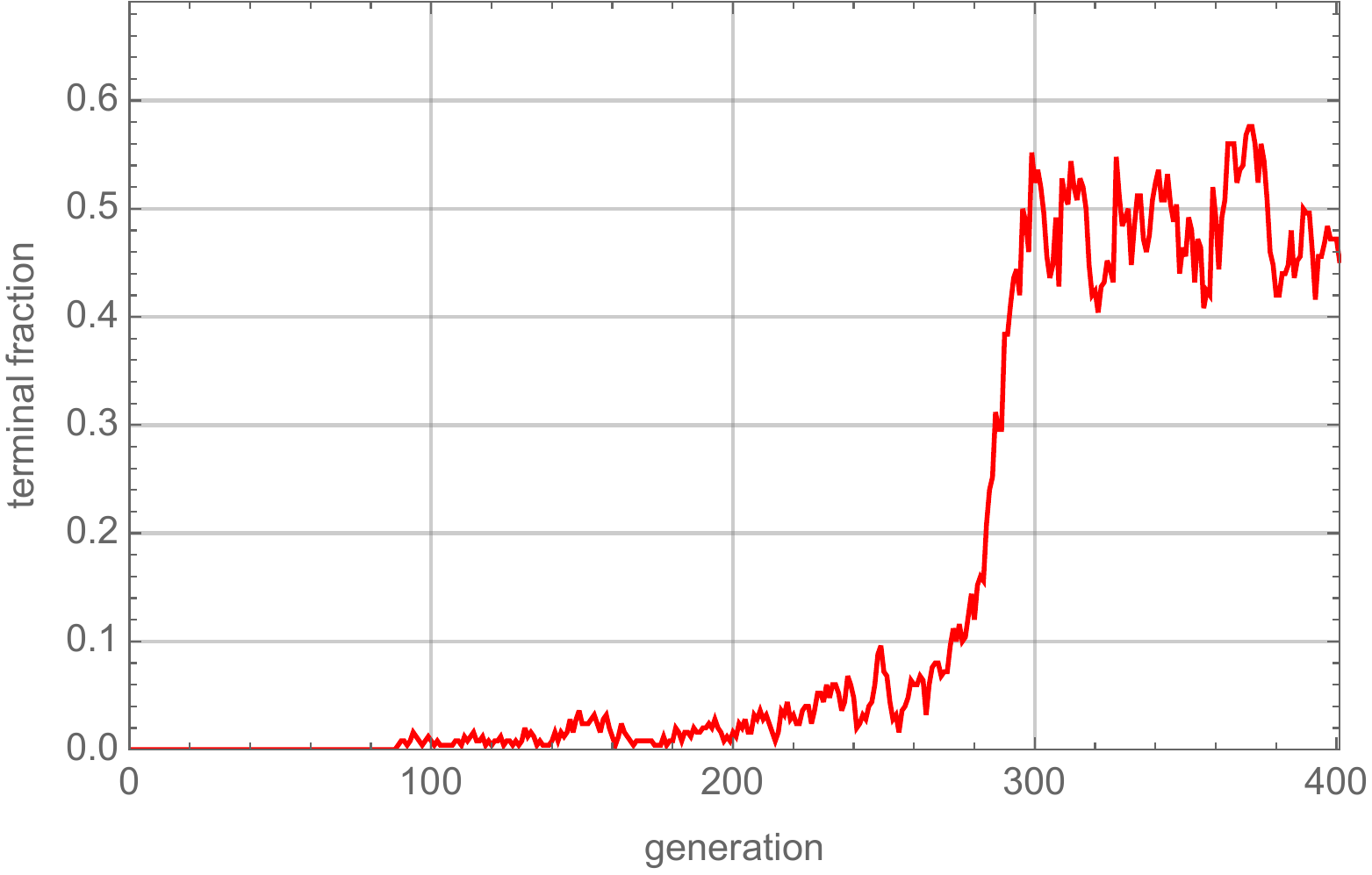}}
     \caption{Performance measures for a typical GA initialisation on the triple trilinear.}
     \label{figExampleTT}
\end{figure}

\begin{figure}[h]
     \centering
    {\includegraphics[width=0.45\textwidth]{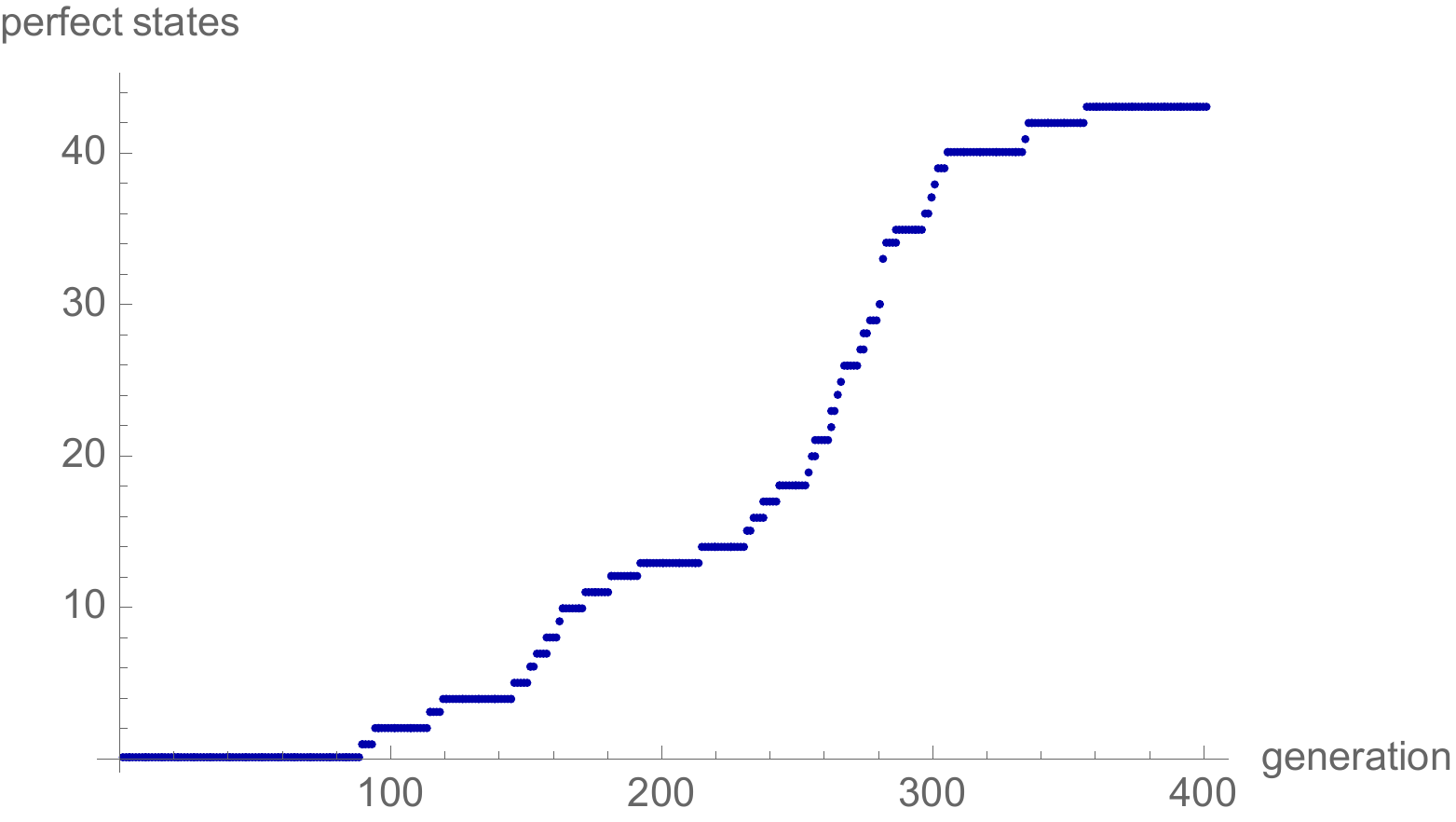}}
       \caption{Saturation of the number of perfect states found in a typical GA run on the triple trilinear CY manifold. }
     \label{fig:TTPerfectStates}
\end{figure}

\noindent Figure~\ref{figExampleTT} shows a typical evolution of a population of models on the triple trilinear manifold. By generation $300$ about half of the population corresponds to perfect models. For this run, a total of $400\times 250$ states have been visited, many of them multiple times. Of these $15,377$ correspond to perfect models, with $43$ being distinct. After eliminating redundancies, $41$ non-equivalent models remain. \\[2mm]
The size of the search space is 
\[
 8^{21}\simeq 10^{19}\; .
\] 
which is orders of magnitude larger than the bi-cubic space. The number of states visited during the above run, namely $100,000$ states (not necessarily distinct), represents a minuscule fraction of the space. Nevertheless, it is remarkable that $43$ perfect states were found, and these perfect states lead to $41$ inequivalent models. The total number of perfect states found after $n$ generations as a function of $n$ is plotted in Figure~\ref{fig:TTPerfectStates}, which suggests that the number of generations, $400$, is suitably chosen given the other hyper-parameters. This continued efficacy despite the much larger search space suggests that the difficulty of finding a solution is increasing only polynomially for the GA as one might have hoped. 

\subsection{GA search on the triple trilinear}\label{sec:scanTT}
The environment being much larger than in the case of the bicubic manifold, we have not attempted to reach the same degree of comprehensiveness as before. 
\begin{figure}[h]
     \centering
    {\includegraphics[width=0.5\textwidth]{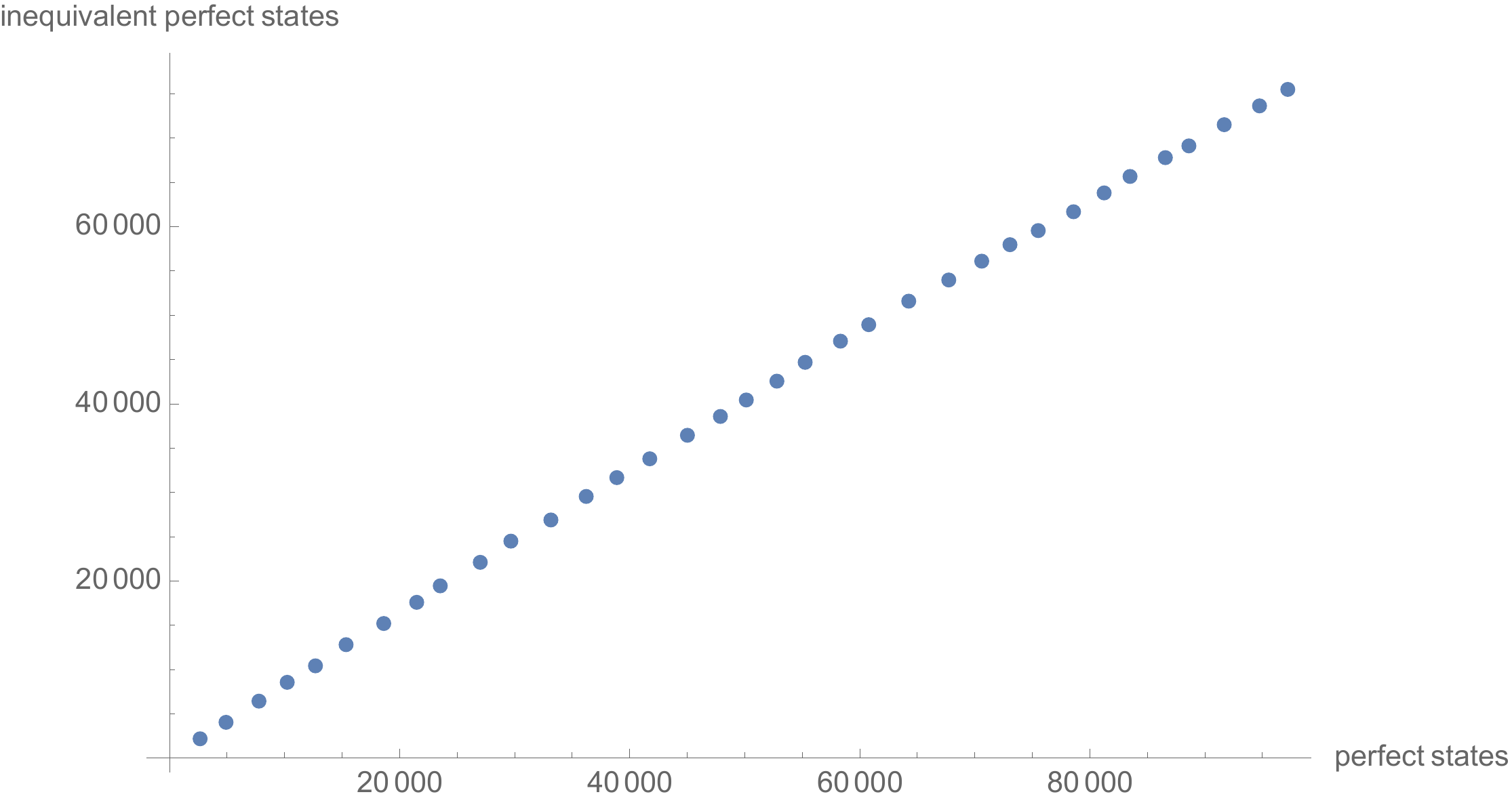}}
       \caption{GA search results on the triple trilinear manifold. The creation of each point has taken 10 core hours and corresponds to $10^8$ visited states. The number of inequivalent perfect states has not saturated after $35\cdot 10^8$ visited states.}
     \label{fig:TTGASaturation}
\end{figure}

\noindent After running for $15$ core days, the GA found $\sim 100,000$ perfect states, three quarters of which are inequivalent. Figure~\ref{fig:TTGASaturation} shows that the number of inequivalent states is far from saturation at the end of the run, suggesting that many more other inequivalent perfect states are still to be found within the environment.

\section{Conclusion}

In this paper we have explored the efficacy of genetic algorithms (GAs) as a search method for probing the landscape of four-dimensional ${\cal N}=1$ heterotic compactifications. GAs turn out to be very successful for engineering the compactification data and for finding perfect models, that is, models which resemble the Standard Model. The spaces of models scrutinised here were extremely large ($\sim\!10^{12}$ in the case of the bicubic manifold and $\sim\!10^{19}$ for the triple trilinear manifold). In the case of the bicubic manifold a systematic scan, although feasible, would be extremely time consuming, while for the triple trilinear manifold, a systematic scan is practically unfeasible.  Despite this, the GA was able to find tens of models within the first few minutes of running on a single machine. The method is not only efficient, but also highly parallelisable. \\[2mm]
After 10 days of running, the GA accomplished a high degree of comprehensiveness in finding all the perfect models available on the bicubic manifold. Indirect evidence in support of this claim was obtained by exploiting the degeneracy of the environment, as well as by comparison of the results of the GA search with those obtained through reinforcement learning (RL). Although the two methods are very different, they lead to datasets of inequivalent models that have an overlap of over $90\%$. This suggests that the details of the optimisation process are not really essential once the processes begin to saturate, provided that they share the same incentives.\\[2mm]
Our comparative analysis of the two search methods revealed that GA is, overall, more efficient by about an order of magnitude in finding perfect states than RL. However, GA is also more prone to finding models that are related by redundant permutations, a feature which is likely related to the crossover mechanism. In principle techniques such as crowding penalties (with the redundancy incorporated) might improve this aspect. An important factor contributing to the lower output of the RL search was the fact that certain perfect states had very large basins of attraction and consequently were found many times, while other perfect states had small basins of attraction and needed many trial episodes to be found. \\[2mm] 
The combined RL and GA searches on the bicubic manifold led to some $700$ inequivalent perfect models. Only a tenth of these correspond to monad bundles of genuine $({\rm rk}(B),{\rm rk}(C))=(6,2)$ type with the correct cohomology dimensions. Further stability checks reduce the number of viable models by another factor of $2$ -- the corresponding monad bundles have been tabulated in Appendix~A. Unfortunately these checks had to be performed at the end of the search due to the limitations imposed by cohomology computations. However, it may be the case that a quick method for computing monad bundle cohomology can be developed, for instance in the form of an analytic formula similar to those found for line bundle sums. If possible, this would facilitate the implementation of further checks at the search stage, which would significantly improve the performance of both GAs and RL.  \\[2mm] 
Improving the performance of the search methods is of particular importance for large environments, such as that corresponding to the triple trilinear manifold. In the absence of more stringent notion of perfect model, achieving any degree of comprehensiveness is difficult in such large environments; consequently, partial searches such as the one presented in Section~\ref{sec:scanTT} may miss many relevant models. 

\noindent Looking more broadly, one is irresistably drawn to the question of whether the optimisation processes that we have been exploring here might have a bearing on physics in the early Universe, in the sense that a qualitatively similar kind of selection may have happened involving, for instance, topological and small-instanton transitions between different vacua. Of course it is not known what (if any) incentives might favour the physics that we observe in our world, although there have over the years been varyingly speculative suggestions (and we do not need to add to the list). But if this is the case and if the idea of a multiverse is correct, then the results that we observe with GA and RL searches do at least support the idea  that selection mechanisms can successfully operate within the vast search space, and indeed that they do so in a quite convergent way. This suggests the possibility of alternative pictures, quite distinct from that of a randomly populated multiverse, in which the properties of this world would be more common than one would naively expect.

\section*{Acknowledgements}
A.~C. is supported by a Stephen Hawking Fellowship, EPSRC grant EP/T016280/1, and T.~R.~H is supported by an STFC studentship. We would like to thank Sven Krippendorf for suggesting GAs as a complementary search method to RL.

\newpage	
\appendix
\section{Monad bundles on the bicubic}\label{appbicubic62}

The table below contains the $28$ monad bundles of genuine type $(r_B,r_C)=(6,2)$ obtained on the bicubic manifold $X$, leading to models with $27$ families and no anti-families. For these bundles equivariance with respect to the $\IZ_3\times\IZ_3$ action on the manifold has been checked at the level of index divisibility for the line bundles in $B$ and $C$. A number of necessary stability checks have been carried out. More precisely, we have found all the line bundles with entries in the range $[-3,3]$ that inject into $V$ and have checked that there exists a non-empty locus in the K\"ahler cone where all these line bundles have a negative slope, thus leaving $V$ slope poly-stable. \\[2mm] 
In order to save space we use the notation $\cO(k_1,k_2)$ instead of $\cO_X(k_1,k_2)$. All the line bundles are to be understood on the bicubic threefold.

\begin{center}
\begin{longtable}{||c|c||} 
 \hline
\varstr{14pt}{7pt}$ B$ & $C$\\
 \hline\hline
\varstr{14pt}{7pt} $~~~~~~~~\mathcal{O}$(0,1)$\oplus\mathcal{O}$(0,1)$\oplus\mathcal{O}$(0,1)$\oplus\mathcal{O}$(1,-1)$\oplus\mathcal{O}$(1,1)$\oplus\mathcal{O}$(1,1)~~~~~~~&$~~~~\mathcal{O}$(1,2)$\oplus\mathcal{O}$(2,2)~~~~\\ \hline
\varstr{14pt}{7pt} $\mathcal{O}$(-1,1)$\oplus\mathcal{O}$(0,1)$\oplus\mathcal{O}$(0,1)$\oplus\mathcal{O}$(0,1)$\oplus\mathcal{O}$(2,-1)$\oplus\mathcal{O}$(2,1)&$\mathcal{O}$(1,2)$\oplus\mathcal{O}$(2,2)\\ \hline
\varstr{14pt}{7pt} $\mathcal{O}$(0,1)$\oplus\mathcal{O}$(0,1)$\oplus\mathcal{O}$(0,1)$\oplus\mathcal{O}$(1,-1)$\oplus\mathcal{O}$(1,1)$\oplus\mathcal{O}$(1,2)&$\mathcal{O}$(1,4)$\oplus\mathcal{O}$(2,1)\\ \hline
\varstr{14pt}{7pt} $\mathcal{O}$(0,1)$\oplus\mathcal{O}$(0,1)$\oplus\mathcal{O}$(0,1)$\oplus\mathcal{O}$(1,0)$\oplus\mathcal{O}$(1,0)$\oplus\mathcal{O}$(1,0)&$\mathcal{O}$(1,1)$\oplus\mathcal{O}$(2,2)\\ \hline
\varstr{14pt}{7pt} $\mathcal{O}$(-1,1)$\oplus\mathcal{O}$(-1,1)$\oplus\mathcal{O}$(1,1)$\oplus\mathcal{O}$(1,1)$\oplus\mathcal{O}$(1,1)$\oplus\mathcal{O}$(2,-1)&$\mathcal{O}$(1,2)$\oplus\mathcal{O}$(2,2)\\ \hline
\varstr{14pt}{7pt} $\mathcal{O}$(0,1)$\oplus\mathcal{O}$(0,1)$\oplus\mathcal{O}$(0,1)$\oplus\mathcal{O}$(1,-2)$\oplus\mathcal{O}$(1,2)$\oplus\mathcal{O}$(2,1)&$\mathcal{O}$(2,2)$\oplus\mathcal{O}$(2,2)\\ \hline
\varstr{14pt}{7pt} $\mathcal{O}$(-2,1)$\oplus\mathcal{O}$(1,0)$\oplus\mathcal{O}$(1,0)$\oplus\mathcal{O}$(1,0)$\oplus\mathcal{O}$(1,2)$\oplus\mathcal{O}$(2,1)&$\mathcal{O}$(2,2)$\oplus\mathcal{O}$(2,2)\\ \hline
\varstr{14pt}{7pt} $\mathcal{O}$(-2,2)$\oplus\mathcal{O}$(1,0)$\oplus\mathcal{O}$(1,0)$\oplus\mathcal{O}$(1,0)$\oplus\mathcal{O}$(1,1)$\oplus\mathcal{O}$(1,1)&$\mathcal{O}$(1,2)$\oplus\mathcal{O}$(2,2)\\ \hline
\varstr{14pt}{7pt} $\mathcal{O}$(-1,1)$\oplus\mathcal{O}$(-1,2)$\oplus\mathcal{O}$(1,0)$\oplus\mathcal{O}$(1,0)$\oplus\mathcal{O}$(1,0)$\oplus\mathcal{O}$(2,1)&$\mathcal{O}$(1,2)$\oplus\mathcal{O}$(2,2)\\ \hline
\varstr{14pt}{7pt} $\mathcal{O}$(-1,2)$\oplus\mathcal{O}$(0,1)$\oplus\mathcal{O}$(0,1)$\oplus\mathcal{O}$(0,1)$\oplus\mathcal{O}$(2,-2)$\oplus\mathcal{O}$(2,1)&$\mathcal{O}$(1,2)$\oplus\mathcal{O}$(2,2)\\ \hline
\varstr{14pt}{7pt} $\mathcal{O}$(0,2)$\oplus\mathcal{O}$(0,2)$\oplus\mathcal{O}$(0,2)$\oplus\mathcal{O}$(1,-2)$\oplus\mathcal{O}$(1,-1)$\oplus\mathcal{O}$(1,1)&$\mathcal{O}$(1,2)$\oplus\mathcal{O}$(2,2)\\ \hline
\varstr{14pt}{7pt} $\mathcal{O}$(-1,2)$\oplus\mathcal{O}$(-1,2)$\oplus\mathcal{O}$(1,-1)$\oplus\mathcal{O}$(1,-1)$\oplus\mathcal{O}$(1,1)$\oplus\mathcal{O}$(2,1)&$\mathcal{O}$(1,2)$\oplus\mathcal{O}$(2,2)\\ \hline
\varstr{14pt}{7pt} $\mathcal{O}$(-1,1)$\oplus\mathcal{O}$(-1,1)$\oplus\mathcal{O}$(1,1)$\oplus\mathcal{O}$(1,1)$\oplus\mathcal{O}$(1,1)$\oplus\mathcal{O}$(2,-1)&$\mathcal{O}$(1,2)$\oplus\mathcal{O}$(2,2)\\ \hline
\varstr{14pt}{7pt} $\mathcal{O}$(-1,1)$\oplus\mathcal{O}$(0,1)$\oplus\mathcal{O}$(0,1)$\oplus\mathcal{O}$(0,1)$\oplus\mathcal{O}$(2,-1)$\oplus\mathcal{O}$(2,1)&$\mathcal{O}$(1,2)$\oplus\mathcal{O}$(2,2)\\ \hline
\varstr{14pt}{7pt} $\mathcal{O}$(-1,1)$\oplus\mathcal{O}$(-1,1)$\oplus\mathcal{O}$(-1,1)$\oplus\mathcal{O}$(2,0)$\oplus\mathcal{O}$(2,0)$\oplus\mathcal{O}$(2,0)&$\mathcal{O}$(1,1)$\oplus\mathcal{O}$(2,2)\\ \hline
\varstr{14pt}{7pt} $\mathcal{O}$(-1,1)$\oplus\mathcal{O}$(-1,1)$\oplus\mathcal{O}$(1,1)$\oplus\mathcal{O}$(1,1)$\oplus\mathcal{O}$(1,2)$\oplus\mathcal{O}$(2,-1)&$\mathcal{O}$(1,4)$\oplus\mathcal{O}$(2,1)\\ \hline
\varstr{14pt}{7pt} $\mathcal{O}$(-2,2)$\oplus\mathcal{O}$(-1,1)$\oplus\mathcal{O}$(1,1)$\oplus\mathcal{O}$(1,4)$\oplus\mathcal{O}$(2,-1)$\oplus\mathcal{O}$(2,-1)&$\mathcal{O}$(1,5)$\oplus\mathcal{O}$(2,1)\\ \hline
\varstr{14pt}{7pt} $\mathcal{O}$(-2,1)$\oplus\mathcal{O}$(1,0)$\oplus\mathcal{O}$(1,0)$\oplus\mathcal{O}$(1,0)$\oplus\mathcal{O}$(1,2)$\oplus\mathcal{O}$(1,4)&$\mathcal{O}$(1,5)$\oplus\mathcal{O}$(2,2)\\ \hline
\varstr{14pt}{7pt} $\mathcal{O}$(-1,1)$\oplus\mathcal{O}$(-1,2)$\oplus\mathcal{O}$(1,0)$\oplus\mathcal{O}$(1,0)$\oplus\mathcal{O}$(1,0)$\oplus\mathcal{O}$(1,4)&$\mathcal{O}$(1,2)$\oplus\mathcal{O}$(1,5)\\ \hline
\varstr{14pt}{7pt} $\mathcal{O}$(-1,2)$\oplus\mathcal{O}$(-1,2)$\oplus\mathcal{O}$(1,-1)$\oplus\mathcal{O}$(1,-1)$\oplus\mathcal{O}$(1,1)$\oplus\mathcal{O}$(1,4)&$\mathcal{O}$(1,2)$\oplus\mathcal{O}$(1,5)\\ \hline
\varstr{14pt}{7pt} $\mathcal{O}$(0,1)$\oplus\mathcal{O}$(0,1)$\oplus\mathcal{O}$(0,1)$\oplus\mathcal{O}$(1,-2)$\oplus\mathcal{O}$(1,2)$\oplus\mathcal{O}$(1,4)&$\mathcal{O}$(1,5)$\oplus\mathcal{O}$(2,2)\\ \hline
\varstr{14pt}{7pt} $\mathcal{O}$(0,1)$\oplus\mathcal{O}$(0,1)$\oplus\mathcal{O}$(0,1)$\oplus\mathcal{O}$(1,-2)$\oplus\mathcal{O}$(1,4)$\oplus\mathcal{O}$(4,1)&$\mathcal{O}$(1,5)$\oplus\mathcal{O}$(5,1)\\ \hline
\varstr{14pt}{7pt} $\mathcal{O}$(-1,2)$\oplus\mathcal{O}$(-1,2)$\oplus\mathcal{O}$(1,0)$\oplus\mathcal{O}$(1,0)$\oplus\mathcal{O}$(1,0)$\oplus\mathcal{O}$(2,-1)&$\mathcal{O}$(1,2)$\oplus\mathcal{O}$(2,1)\\ \hline
\varstr{14pt}{7pt} $\mathcal{O}$(-2,2)$\oplus\mathcal{O}$(1,0)$\oplus\mathcal{O}$(1,0)$\oplus\mathcal{O}$(1,0)$\oplus\mathcal{O}$(1,1)$\oplus\mathcal{O}$(1,2)&$\mathcal{O}$(1,4)$\oplus\mathcal{O}$(2,1)\\ \hline
\varstr{14pt}{7pt} $\mathcal{O}$(-1,1)$\oplus\mathcal{O}$(-1,1)$\oplus\mathcal{O}$(1,1)$\oplus\mathcal{O}$(1,1)$\oplus\mathcal{O}$(1,1)$\oplus\mathcal{O}$(1,1)&$\mathcal{O}$(1,2)$\oplus\mathcal{O}$(1,4)\\ \hline
\varstr{14pt}{7pt} $\mathcal{O}$(-1,1)$\oplus\mathcal{O}$(-1,2)$\oplus\mathcal{O}$(1,0)$\oplus\mathcal{O}$(1,0)$\oplus\mathcal{O}$(1,0)$\oplus\mathcal{O}$(1,2)&$\mathcal{O}$(1,1)$\oplus\mathcal{O}$(1,4)\\ \hline
\varstr{14pt}{7pt} $\mathcal{O}$(-2,1)$\oplus\mathcal{O}$(1,0)$\oplus\mathcal{O}$(1,0)$\oplus\mathcal{O}$(1,0)$\oplus\mathcal{O}$(1,4)$\oplus\mathcal{O}$(4,1)&$\mathcal{O}$(1,5)$\oplus\mathcal{O}$(5,1)\\ \hline
\varstr{14pt}{7pt} $\mathcal{O}$(-1,2)$\oplus\mathcal{O}$(0,1)$\oplus\mathcal{O}$(0,1)$\oplus\mathcal{O}$(0,1)$\oplus\mathcal{O}$(2,-1)$\oplus\mathcal{O}$(2,-1)&$\mathcal{O}$(1,2)$\oplus\mathcal{O}$(2,1)\\ \hline
\end{longtable}
\end{center}

\providecommand{\href}[2]{#2}\begingroup\raggedright\endgroup

\end{document}